\documentclass[preprint, amsmath, amssymb, aps, prl, superscriptaddress]{revtex4-2}

\usepackage{mathtools}
\usepackage{graphicx}
\usepackage{dcolumn}
\usepackage{bm}
\usepackage{xcolor}
\usepackage{ragged2e}
\usepackage{float}
\usepackage{soul}

\setstcolor{red}

\begin{document}

\title{Spin wavepackets in the Kagome ferromagnet Fe$_3$Sn$_2$: propagation and precursors}

\author{Changmin Lee}
\thanks{These authors contributed equally.}
\affiliation{Materials Science Division, Lawrence Berkeley National Laboratory, Berkeley, CA 94720, USA}

\author{Yue Sun}
\thanks{These authors contributed equally.}
\affiliation{Materials Science Division, Lawrence Berkeley National Laboratory, Berkeley, CA 94720, USA}
\affiliation{Department of Physics, University of California at Berkeley, Berkeley, CA 94720, USA}

\author{Linda Ye}
\thanks{Present Address: Department of Applied Physics, Stanford University, Stanford, CA 94305, USA}
\affiliation{Department of Physics, Massachusetts Institute of Technology, Cambridge, MA 02139, USA}

\author{Sumedh Rathi}
\affiliation{Materials Science Division, Lawrence Berkeley National Laboratory, Berkeley, CA 94720, USA}
\affiliation{Department of Physics, University of California at Berkeley, Berkeley, CA 94720, USA}

\author{Kevin Wang}
\affiliation{Materials Science Division, Lawrence Berkeley National Laboratory, Berkeley, CA 94720, USA}
\affiliation{Department of Physics, University of California at Berkeley, Berkeley, CA 94720, USA}

\author{Yuan-Ming Lu}
\affiliation{Department of Physics, The Ohio State University, Columbus, OH 43210, USA}

\author{Joel Moore}
\affiliation{Materials Science Division, Lawrence Berkeley National Laboratory, Berkeley, CA 94720, USA}
\affiliation{Department of Physics, University of California at Berkeley, Berkeley, CA 94720, USA}

\author{Joseph G. Checkelsky}
\affiliation{Department of Physics, Massachusetts Institute of Technology, Cambridge, MA 02139, USA}

\author{Joseph Orenstein}
\email{jworenstein@lbl.gov}
\affiliation{Materials Science Division, Lawrence Berkeley National Laboratory, Berkeley, CA 94720, USA}
\affiliation{Department of Physics, University of California at Berkeley, Berkeley, CA 94720, USA}

\date{\today}
\begin{abstract}
The propagation of spin waves in magnetically ordered systems has emerged as a potential means to shuttle quantum information over large distances. Conventionally, the arrival time of a spin wavepacket at a distance, $d$, is assumed to be determined by its group velocity, $v_g$. He we report time-resolved optical measurements of wavepacket propagation in the Kagome ferromagnet Fe$_3$Sn$_2$ that demonstrate the arrival of spin information at times significantly less than $d/v_g$. We show that this spin wave ``precursor" originates from the interaction of light with the unusual spectrum of magnetostatic modes in Fe$_3$Sn$_2$.  Related effects may have far-reaching consequences toward realizing long-range, ultrafast spin wave transport in both ferromagnetic and antiferromagnetic systems. 
\end{abstract}

\maketitle

\section{Introduction}
Harnessing electron spin is one of the central goals of condensed matter physics. A particularly exciting direction is the coupling of spin to charge and lattice degrees of freedom to provide interconnections in hybrid quantum systems. To this end, it is essential to understand and control the generation, propagation, and detection of spin information. Recent progress in magnetically ordered systems has shown promise in using propagating spin waves -- collective excitations of the electron spins -- to transport information over large distances \cite{Kajiwara2010,Cornelissen2015,Lebrun2018,Chumak2015,Pirro2021}. Increasingly, attention has focused on quasi-two dimensional (2D) layered magnets in which spins within each plane are parallel but the interplane order can be ferromagnetic \cite{Huang2017}, antiferromagnetic \cite{Lee2016}, or even helical \cite{Song2022}.

An important subset of such systems is ``easy-plane" magnets in which spins are oriented parallel to the planes but without a preferred direction within the plane. As a result of the symmetry with respect to in-plane spin rotation, the out-of-plane, or $z$ component of the magnetization, $M_z$, is a conserved quantity. Theoretically, $M_z$ can exhibit ballistic, diffusive, hydrodynamic, or even superfluid regimes of transport {\cite{Sonin2010}}.  However, in real 2D easy plane magnets this rotational symmetry is broken, although weakly, by the anisotropy of the underlying lattice. This fact has driven theoretical studies of the consequences of rotational symmetry breaking and approaches to mitigating its effects \cite{Shen2021,Qaiumzadeh2017}.

At low temperatures, Fe$_3$Sn$_2$ exemplifies an easy-plane system of the class introduced above, in which the spins experience weak anisotropy resulting from the discrete 3-fold rotational symmetry of the rhombohedrally stacked Kagome lattice \cite{LeCaer1978,Ye2018,Kumar2019}. In this work we study the propagation of spin wavepackets in Fe$_3$Sn$_2$, using temporal and spatially resolved optical techniques to probe their amplitude, frequency, and velocity. In our pump/probe measurement scheme, the pump pulse excites a spin wavepacket whose propagation is detected by a time-delayed and spatially-separated probe pulse through the magneto-optic Kerr effect (MOKE) \cite{Hiebert1997,Acremann2000} or optical birefringence \cite{Kimel2004}. The range of wavevectors that comprise the spin wavepacket is determined by the Fourier transform of the real space excitation density, which is typically Gaussian.  

Since the size of the focused laser spot is diffraction limited, the excited wavevectors are typically within the range of inverse micrometers ($\mu \textrm{m}^{-1}$). In this long wavelength regime, the propagation of spin is dominated by magnetic dipole interactions, drastically altering the properties that arise from short-ranged exchange interactions alone \cite{Damon1961,Stancil2009}. Excitations in this regime are referred to as magnetostatic spin waves (MSWs) although they are fully dynamic; the term arises because their dispersion relations can be obtained within the magnetostatic approximation, $\nabla\times \bf{H}=0$, which is valid because spin wave velocities are much smaller than the speed of light.  

Given the long-range nature of the dipole interaction, MSWs are particularly sensitive to both the shape of the medium and magnetic anisotropy. Damon and Eshbach (DE) \cite{Damon1961} obtained MSW dispersion relations for a magnetic slab with uniaxial anisotropy, associated with either an easy axis or applied magnetic field. However, as we demonstrate below, spin wavepacket propagation in Fe$_3$Sn$_2$ shows novel properties that cannot be described by the DE relations, including the remarkable observation that spin excitations can be detected remotely at a time much shorter than would be inferred from the spin wave velocity. In the theoretical component our study, we use both analytical calculations and numerical modeling to show that these effects are accounted by extending the DE formalism to the easy plane systems of current interest. Although the theory presented below assumes ferromagnetic order between the planes, as in Fe$_3$Sn$_2$, it applies to antiferromagnetic order as well.  For example, the antiferromagnetic version of the theory provides a quantitative explanation for the recently discovered surprising properties of spin wavepacket propagation in the 2D van der Waals antiferromagnet CrSBr \cite{Bae2022}. 

\section{Results}
\subsection{Magnetic field dependence of spin wave frequency}

Prior to measurements of spin transport, the anisotropy parameters of Fe$_3$Sn$_2$ were determined using the time-resolved magneto-optic Kerr effect (TR-MOKE). In this method, a short ($\sim$100 fs) duration  laser pulse induces a transient misalignment between the magnetization, $\bf{M}$, and the effective anisotropy field, $\bf{H}_\text{eff}$. The resulting torque causes $\bf{M}$ to precess, as illustrated in Fig. 1(a). The precession leads to oscillations of the component of magnetization parallel to the optic axis, $M_z$, which are detected via the polar Kerr effect. \cite{Hiebert1997,Acremann2000,Langner2009}.  

\begin{figure} [ht]
\includegraphics[width=5in]{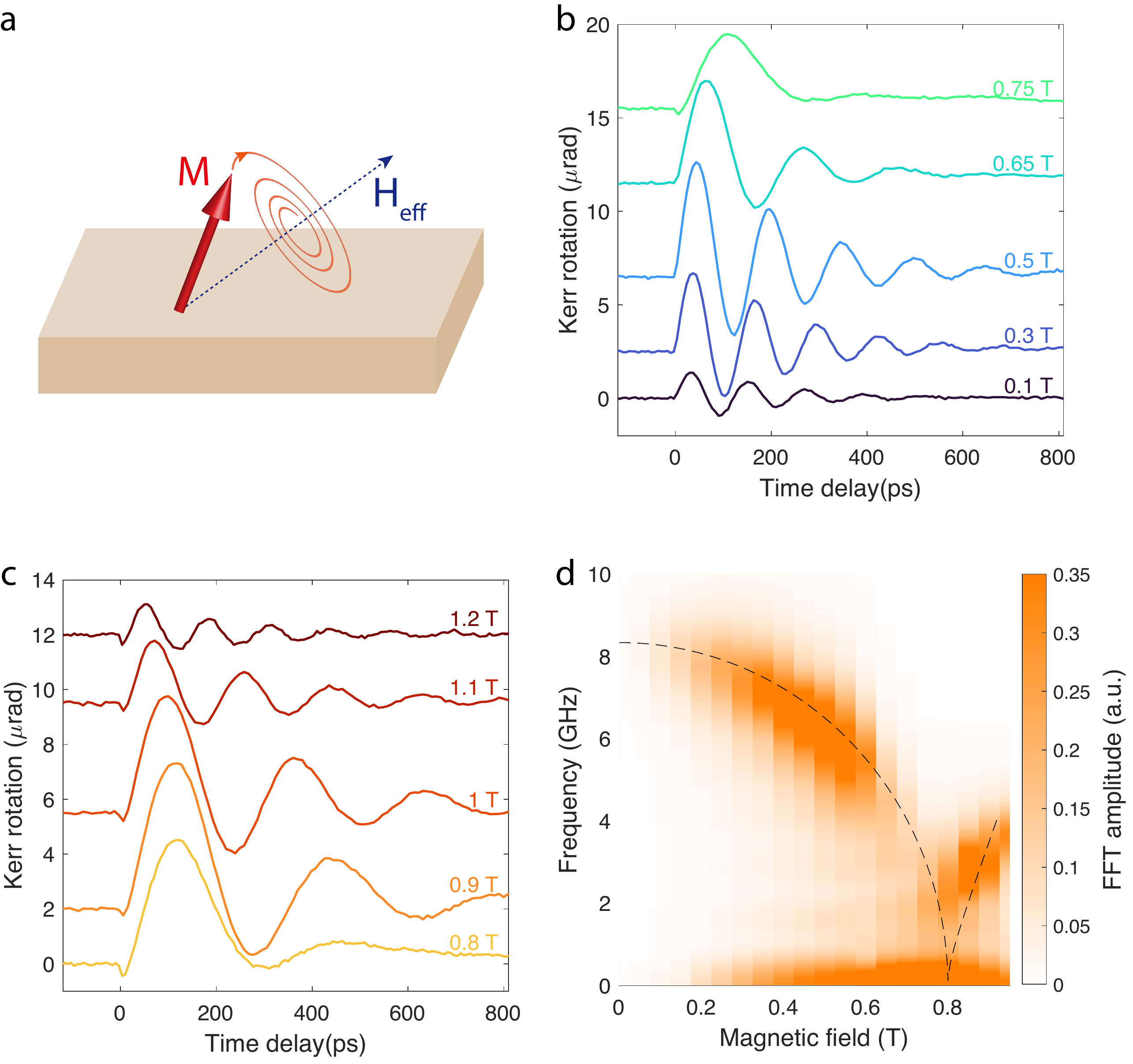}
\caption{\label{fig1} (a) Illustration of magnetization spiraling to align with $H_\text{eff}$. (b,c) Kerr rotation as function of pump-probe delay shown for applied magnetic fields $\leq0.75$ and $> 0.75$ T, respectively. The curves are offset for clarity. (d) Amplitude of Fourier transforms of the time series plotted in (b) and (c) shown in the frequency-field plane. The dashed line indicates the fit to Eq. \ref{eq:FMR_Hdep}. All data were taken at $T = 2.5\ \textrm{K}$}
\end{figure}

Figures \ref{fig1}(b) and \ref{fig1}(c) show oscillations of $M_z$ as detected by the TR-MOKE for several magnetic fields applied in the $z$ direction. Fig. \ref{fig1}(d) displays the Fourier transform of the oscillations in the frequency-magnetic field plane; the dashed line is a fit to a model described below. This dependence of spin wave (SW) frequency on field is characteristic of a ferromagnet whose biaxial anisotropy can be described by the free energy
$F_{A}=(-K_{x}M_{x}^{2}+K_{z}M_{z}^{2})/M_{s}^{2}$, where $K_x$ and $K_z$ are the in- and out-of-plane anisotropy energies ($K_x, K_z > 0$) and $x$ is a preferred magnetization direction within the plane. The origin of the in-plane anisotropy is discussed in Supporting Information Sections I and II, which present a microscopic model whose low-temperature, broken-symmetry phase is described by $F_{A}$ for small fluctuations of the ferromagnetic order parameter. The theoretically predicted dependence of SW frequency on magnetic field, $\nu(H_z)$, for a biaxial ferromagnet is \cite{Suhl1955, Demokritov1989},


\begin{align} \label{eq:FMR_Hdep}
\nu(H_z) =
    \begin{dcases*}
        \frac{\gamma\sqrt{(H_{s}^{2}-H_{z}^{2})K_{x}(K_{x}+K_{z}+2\pi M_{s}^{2})}}{2\pi(K_{x}+K_{z})} & $0\leqslant H_{z} \leqslant H_{s}$ \\
        \frac{\gamma}{2\pi}\sqrt{\left(H_{z}-\frac{2K_{z}}{M_{s}}\right)(H_{z}-H_{s})} & $H_{z}>H_{s}$
    \end{dcases*}
\end{align}
where $H_{s}$ is the saturation field along the $z$ direction, $M_s$ is the saturation magnetization, and $\gamma$ is the gyromagnetic ratio. Eq. \ref{eq:FMR_Hdep} accurately describes the full field dependence of the TR-MOKE frequencies observed in our experiment (see Supporting Information Section V). The fit (dashed line in Fig. \ref{fig1}d) yields $K_x\approx  1.76 \times 10^4 \text{ J/m}^3$ and $K_z\approx 2.26 \times 10^5 \text{ J/m}^3$, consistent with the picture of weak anisotropy within an easy plane. 


\subsection{Detection of spin propagation using scanning TR-MOKE microscopy}

\begin{figure} [ht]
\includegraphics[width=5in]{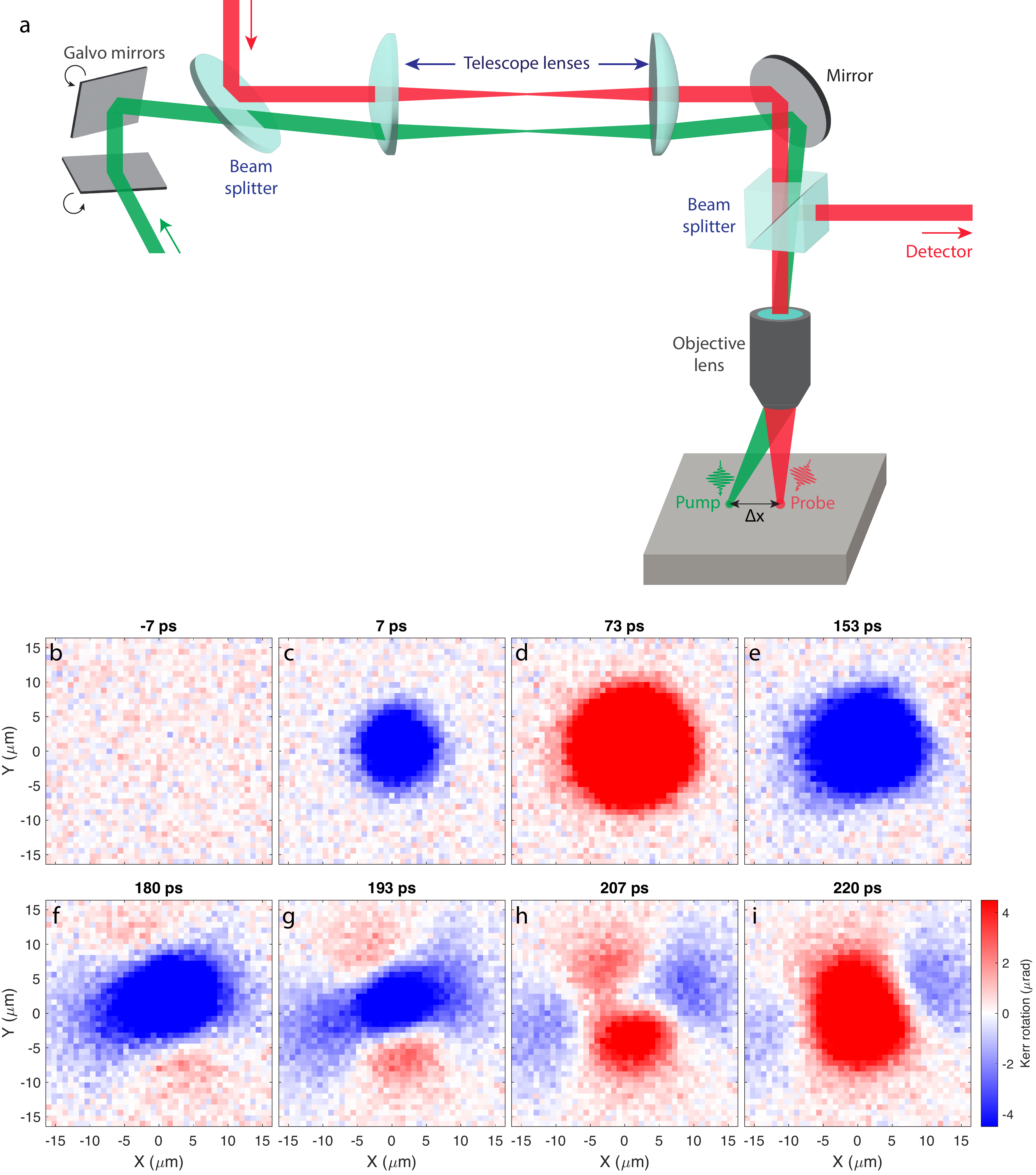}
\caption{\label{fig2} Time-resolved MOKE microscopy (a) Overview of the experimental setup. The 2D galvo mirrors and the $4f$ optical geometry enable scanning of the pump laser beam. (b)-(i) Snapshots of 2D MOKE maps taken at several pump-probe time delays. For time delay $t\geq 180$ ps (f-i), the propagation is clearly anisotropic, with contrasting properties along two principal axis directions.  All measurements were performed at $T = 2.5\ \textrm{K}$ with an out-of-plane field of 0.5 T.}
\end{figure}

We now describe extending time-resolved measurements to the spatial domain. A simplified layout of the setup for TR-MOKE microscopy is shown in Fig. \ref{fig2}(a). The $4f$ optical system equipped with 2-axis galvo-driven mirrors enables continuous scanning of the pump focus in two dimensions while the location of the probe is fixed \cite{Satoh2012}.  Spin waves photoexcited in one location can be probed remotely at a subsequent time, enabling an all-optical ultrafast investigation of SW transport with micron-scale spatial resolution and sub-microradian polarization sensitivity.

Figures \ref{fig2}(b)-(i) show TR-MOKE maps measured at various pump-probe time delays ($\Delta t$) ranging from -7 to +220 ps. Here, the $x$ and $y$ axes refer to the separation between the pump and probe beams. Shortly after photoexcitation ($\Delta t < 100$ ps), the transient changes in magnetization are isotropic (Figs. \ref{fig2}c-e). However, at $\Delta t = 180$  ps, (Fig. \ref{fig2}e) clear evidence of anisotropic propagation is observed, with a stronger MOKE signal along $\sim20^{\circ}/ 200^{\circ}$ with respect to the $x$ axis. At longer times (Figs. \ref{fig2}g-i) the contrasting nature of propagation between the $20^{\circ}/200^{\circ}$ and $110^{\circ}/ 290^{\circ}$ becomes increasingly clear (angular dependence of the anisotropic propagation is further discussed in the Supporting Information Section III).

\begin{figure} [ht]
\includegraphics[width=5in]{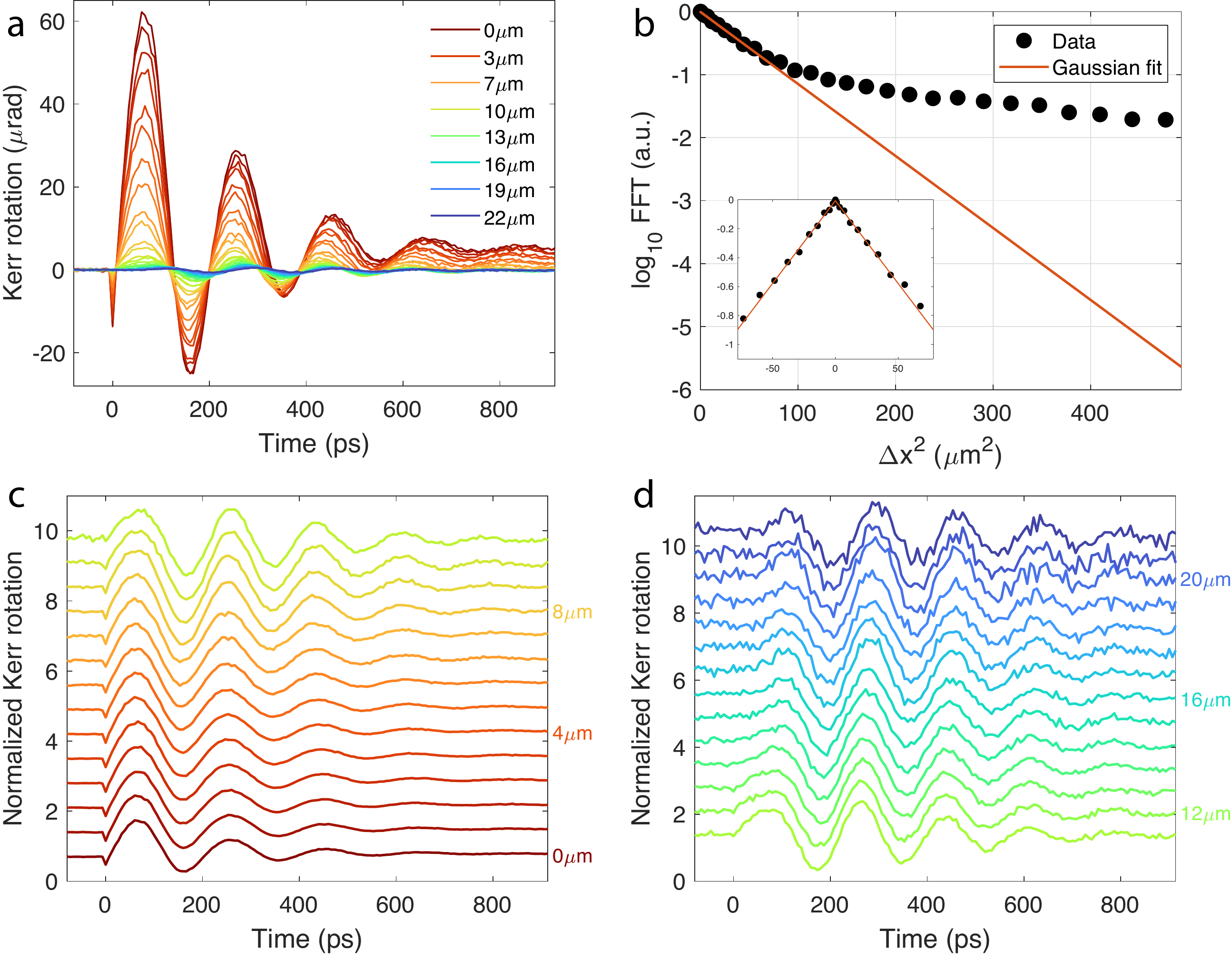}
\caption{\label{fig3}  (a) TR-MOKE traces at different values of spatial separation ($\Delta x$) between the pump and probe beams.  (b) The log of the amplitude of the Fourier transform of the data shown in (a) is plotted vs. $(\Delta x)^2$ (black dots). The red line is the rate of decrease expected in the absence of propagation. (c,d) Normalized TR-MOKE traces at separations for $\Delta x< 10\mu$m and $\Delta x> 10\mu$m, respectively, illustrating the change in the envelope function from exponential to Gaussian. All measurements were taken at $T = 2.5\ \textrm{K}$ under an out-of-plane field of $0.5\ \textrm{T}$.}
\end{figure}

To further characterize spin propagation, we consider the rate of decay of the TR-MOKE oscillations with increasing propagation distance. In Fig. \ref{fig3}(a) we plot TR-MOKE time traces for several values of pump-probe separation along the major propagation axis ($20^{\circ}/200^{\circ})$.  The amplitude at a given separation is determined from the peak value of the Fourier transform of the oscillations.  The log of this amplitude is plotted as a function of $(\Delta x)^2$ as solid circles in Figure \ref{fig3}(b). For small separations the amplitude decreases in proportion to $e^{-(\Delta x/\sigma)^2}$, where $\sigma \sim 6 \mu$m. In this regime, the decay of the amplitude reflects the spatial overlap of pump and probe foci (with full width at half-maximum (FWHM) spot sizes of 6 and 5 $\mu$m, respectively) as would be expected in the absence of propagation. However, for larger $\Delta x$ spin propagation becomes evident; for $\Delta x> 10\mu$m the TR-MOKE amplitude deviates from a Gaussian and at $\Delta x = 20 \mu$m is four orders of magnitude larger than can be accounted for by spatial overlap.

The distinction between the overlap and propagation regimes is also seen by normalizing the TR-MOKE traces to the amplitude at zero separation.  Fig. \ref{fig3}(c) shows the normalized signals for $\Delta x<10 \mu$m, which is in the Gaussian regime of Fig. \ref{fig3}(b). For these separations the envelope of the TR-MOKE oscillations decays monotonically with increasing time, consistent with a simple damped response.  However, at separations greater than 10$\mu$m, shown in Fig. \ref{fig3}(d), the envelope peaks at a nonzero time delay, as expected for a propagating wavepacket. Focusing on the arrival time of the wavepacket at the largest measured separation of 22$\mu$m reveals another surprising feature.  Notice that the first clear indication that spin waves have reached this distance occurs at $\approx$100 ps, from which we estimate an effective velocity of $\approx2\times 10^7$ cm/s. This velocity is six orders of magnitude larger than the group velocity inferred from neutron scattering measurements \cite{Dally2021}. In the following section we show that discrepancy is resolved by considering wavepacket propagation in the magnetostatic regime.

\section{Discussion}
As mentioned in the introduction, spin wavepacket propagation in Fe$_3$Sn$_2$ cannot be described by the DE dispersion relations for either surface or bulk modes. For example the DE surface mode is nonreciprocal, with a single direction of propagation that is reversed for the two opposing surfaces. Instead, we observe reciprocal propagation, that is symmetric with respect to wavevector $\textbf{k}\rightarrow -\textbf{k}$. The volume modes, although reciprocal, propagate only along one axis, whereas we observe propagating modes along two principal axes in the plane. Furthermore, the bidirectional DE volume mode is ``backward moving" in the sense that its phase and group velocities are opposite, whereas we find the two principal axes of propagation exhibit forward and backward modes, respectively. As we show below, extending the DE calculation to nearly easy-plane systems accounts for the novel features observed in our spin transport measurements.

\subsection{Magnetostatic spin waves under biaxial anisotropy}

\begin{figure} [ht!]
\includegraphics[width=5in]{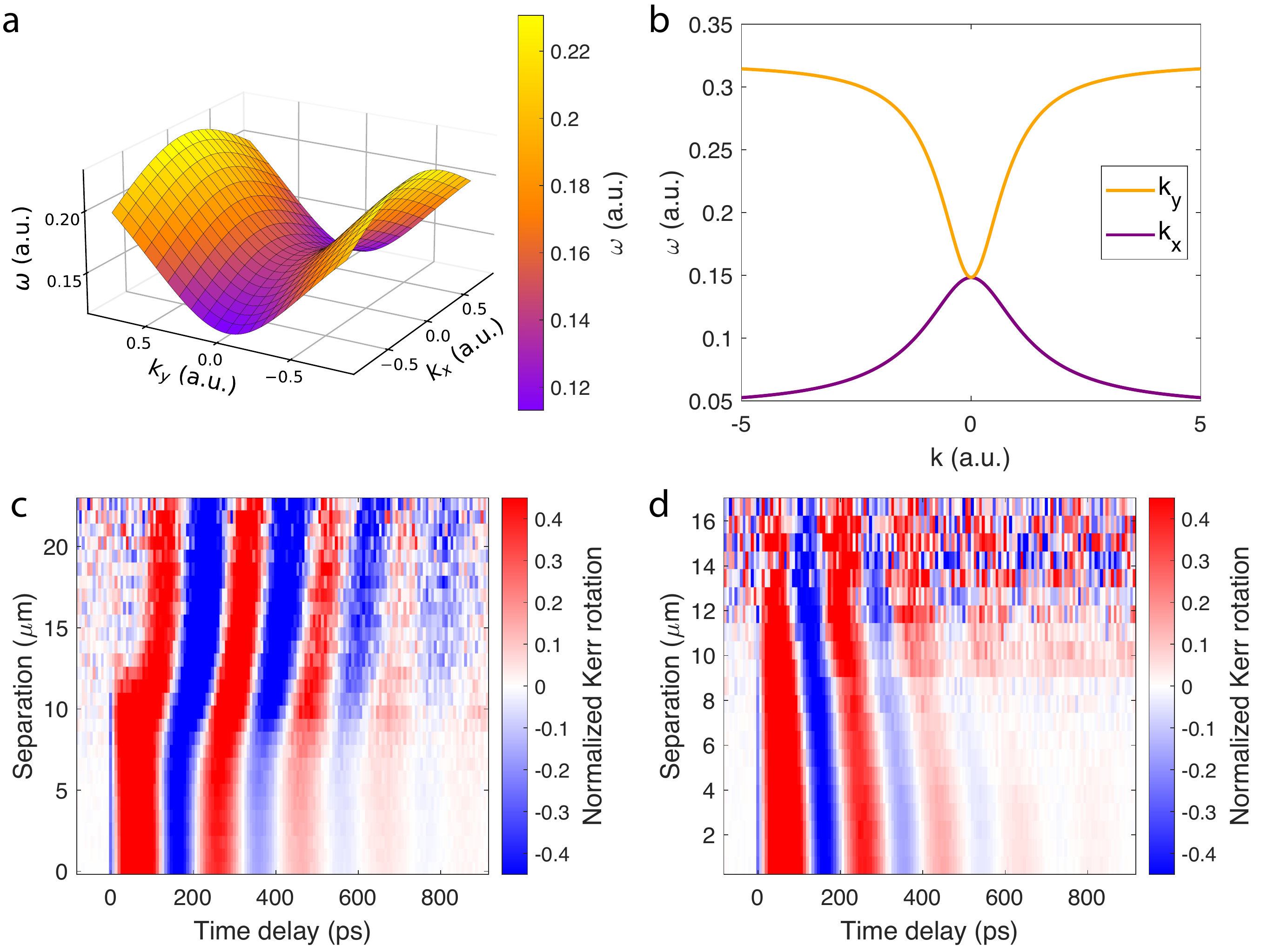}
\caption{\label{fig4} Magnetostatic waves (MSWs) under biaxial anisotropy. (a) Three dimensional representation of the calculated MSW dispersion of Fe$_3$Sn$_2$ as a function of $k_x$ and $k_y$ evaluated at $k_z = 1\ \mu \textrm{m}^{-1}$. A saddle point can be observed at the origin. (b) Frequency-momentum cuts at $k_x = 0$ (purple) and $k_y = 0$ (orange) illustrating forward propagation along $k_y$ (purple) and backward propagation along $k_x$ (orange). (c,d) Plots of the TR-MOKE ampitude in the $\Delta x, t$ plane measured along the two principal axes of propagation show forward and backward propagation, respectively.}
\end{figure}

We consider a geometry with the equilibrium magnetization in the plane and parallel to one of the easy axes \cite{Hurben1996, Hashimoto2017}. Maxwell's equations in the magnetostatic regime, $\nabla \cdot \textbf{B} =\nabla \times \textbf{H} = 0$, together with the Landau-Lifshitz equation,
\begin{equation}
    \frac{\partial \bf{M}}{\partial t}=- \gamma \bf{M}\times H_{eff},
\end{equation}
where $\bf{H_{eff}}$ is the sum of the anisotropy field and the dynamical field $\bf{h}$, form a closed set that yield the normal modes of  magnetization in the long-wavelength regime. To illustrate the resulting MSW dispersion, Fig. \ref{fig4}(a) shows the calculated spin wave frequency in the $k_x,k_y$ plane for fixed $k_z=$1 $\mu m^{-1}$.  Line cuts through this plane defined by $k_x = 0$ (purple) and $k_y = 0$ (orange) plotted in Fig. \ref{fig4}(b) show forward propagation along the $y$ direction and backward along $x$, with a saddle point at $\textbf{q} = 0$. When $K_z > K_x$, as in Fe$_3$Sn$_2$, the velocity is larger along  $k_y$. This dispersion relation was also reproduced through micromagnetic simulations. These predictions are unique to biaxial ferromagnets and clearly distinct from the uniaxial (DE) limit, in which there are no forward propagating reciprocal modes (see Supporting Information Sections IV--VI for the calculations and numerical simulation of the MSW dispersion relations).

The prediction of a saddle dispersion relation was tested by measuring the TR-MOKE oscillations as a function of pump/probe separation along the two principal axes of propagation identified in the maps shown in Fig. \ref{fig2}. The results are presented in Figs. \ref{fig4}(c) and \ref{fig4}(d) as color plots in the time-separation plane. The slope of the lines of constant phase distinguishes forward vs. backward propagating modes. In agreement with our theoretical prediction for the biaxial ferromagnet, modes with wavevector perpendicular to $\bf{M}$ are forward propagating, and backward propagating for wavevectors parallel to $\bf{M}$.

\subsection{Spin wavepacket propagation}

\begin{figure} [ht!]
\includegraphics[width=6.5in]{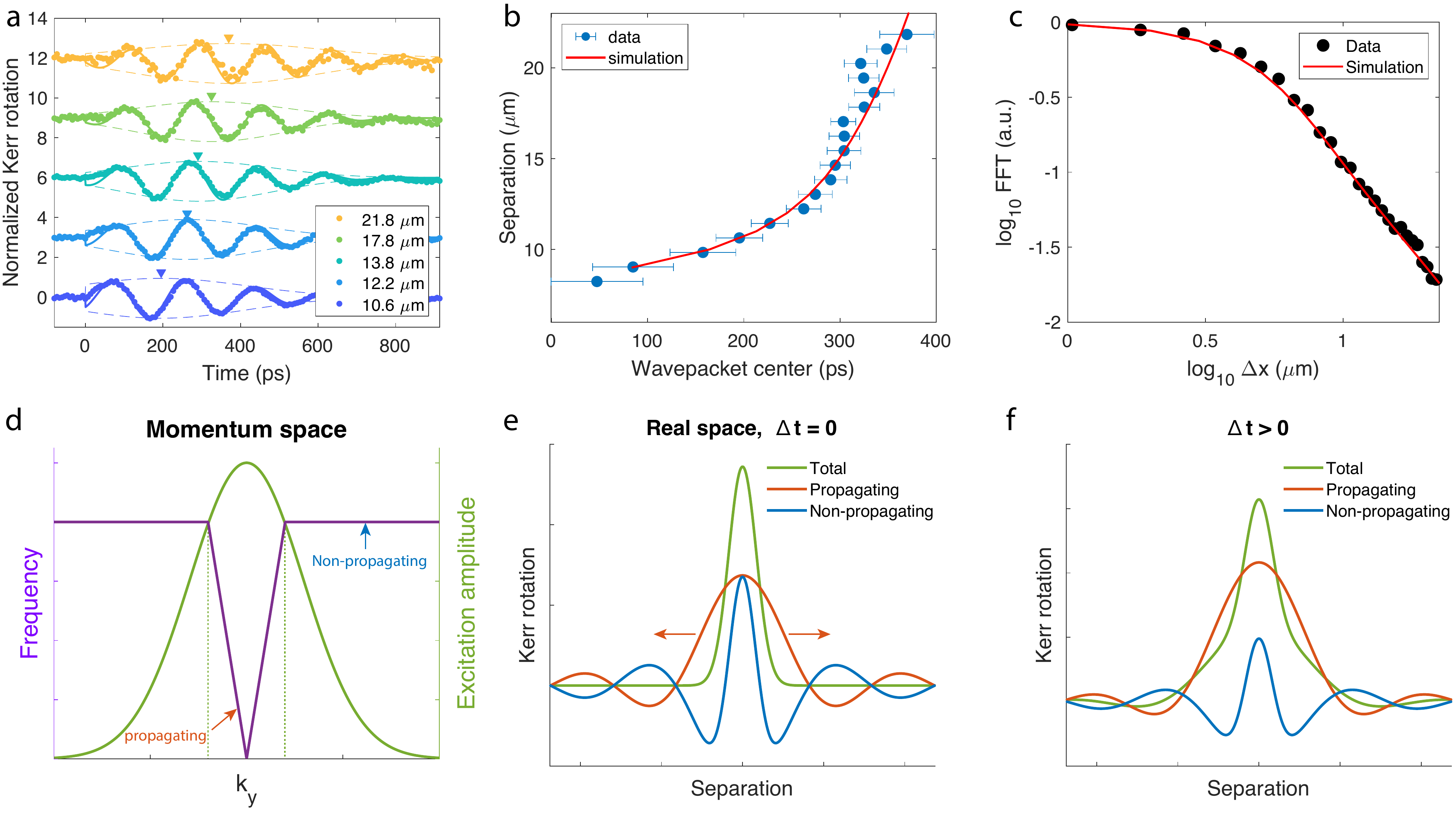}
\caption{\label{fig5} (a) Normalized TR-MOKE amplitude \textit{vs.} $t$ for values of $\Delta x> 10\mu$m with arrowheads indicating the center of the wavepacket. (b) The solid circles show the wavepacket center as a function of time. The red line is a fit based on the calculated MSW dispersion relation. (c) A double logarithmic plot of the amplitude of the wavepacket \textit{vs.} pump-probe separation (solid circles) and the fit (red line) using the same parameters as in (b). (d-f) Illustration of the physical origin of the wavepacket precursor. (d) Shown as a green line is the Gaussian distribution of wavevectors excited by the pump beam. The purple line is an approximation to the MSW dispersion for a value of $k_z$ that is within the range of excited in-plane wavevectors. The regimes with group velocity $v_g>0$ and $v_g=0$ are indicated. (e) Red and blue lines show the contributions to the total Kerr rotation from the propagating and nonpropagating modes, respectively, evaluated at $t=0$. As expected, their sum yields a Gaussian profile corresponding to the initial photoexcited state. (f) As the profile of the propagating modes evolves for $t>0$, the oscillations associated with the nonpropagating modes no longer cancel. The total Kerr rotation shown in green reveals evidence of propagation even at separations greater than $v_g t$.} 
\end{figure}

We turn next to the dynamics of wavepackets whose motion is determined by the dispersion relation illustrated in Fig. \ref{fig4}. The primary goal is to understand how spin information propagates in the MSW regime. Figs. \ref{fig5}(a)-(c) show comparison of experiment and theory for the amplitude and position of the wavepacket.  Fig. \ref{fig5}(a) presents an expanded view of normalized wavepackets measured at several separations larger than $10\mu$m; arrows indicate the time at which the peak amplitude reaches a given distance from the pump. The solid circles in Fig. \ref{fig5}(b) show the displacement of the wavepacket peak as a function of time.  Conventionally, the slope of a fit to these points yields the group velocity, $v_g$. From this perspective, the data are quite puzzling, as $v_g$ appears to increase with time, reaching anomalous value, $\approx 2\times 10^6$ cm/s, much larger that expected for spin waves. Finally, Fig. \ref{fig5}(c) presents a zoomed in view of the wavepacket amplitude vs. separation, now on a double logarithmic plot.

Below we show that the MSW dispersion relation, $\omega(\textbf{k})$, in biaxial magnets successfully explains the anomalous wavepacket propagation in Fe$_3$Sn$_2$.
Crucially for the interpretation of our experiments, photoexcitation launches a coherent spin wavepacket, comprised of a Gaussian distribution of wavevectors that are initially in phase. The time- and position-dependent magnetization detected by TR-MOKE can be calculated using the following relation:
\begin{equation}
    \delta {M_z}(\textbf{r},t) \propto\textrm{Re}\int \textbf{z}\cdot\textbf{m}(\textbf{k})g(\textbf{k})e^{i\textbf{k} \cdot \textbf{r}} e^{-i[\omega(\textbf{k})-i\alpha ]t}d^3k,
\end{equation}
with,
\begin{equation}
    g(\textbf{k})\propto\frac{e^{-\sigma^2(k_x^2+k_y^2)/2} }{ik_z - 1/\delta_p},
\end{equation}
where $g(\textbf{k})$ is the Fourier transform of the initial perturbation generated by the pump pulse and $\textbf{m}(\textbf{k})$ is the normal mode eigenvector. The radius of the focused laser beam, $\sigma$, and the anisotropy parameters are determined from independent measurements.  The only adjustable parameters in the theory are the damping constant, $\alpha$, and the effective penetration depth, $\delta_p$, of the perturbation that induces the subsequent precessional motion. Parameter values, $\delta_p=230$ nm and $\alpha=2.7\times 10^9\ $s$^{-1}$ were chosen to achieve the best fit (solid red line) to the amplitude vs. distance data shown in Fig. \ref{fig5}(c). The same parameters accurately reproduce the anomalous wavepacket position vs. time data as well (red line in Fig. \ref{fig5}(b)), adding additional support for our theoretical model (see Supporting Information Section VII for details).

\subsection{Physical origin of anomalous propagation}

In the previous section we showed that spin wavepacket dynamics in Fe$_3$Sn$_2$ can be quantitatively modeled by the MSW dispersion relations for a biaxial ferromagnet. In this section offer a physical picture that underlies the most puzzling feature of the wavepacket propagation -- apparent velocity in excess of the expected SW velocity. Essentially, the seemingly anomalous behavior is a consequence of a breakdown of the group velocity description that occurs when a dispersion relation is highly structured within the range of wavevectors that comprise the packet. In the following, we show that dynamics in this regime can lead to early arrival times at remote locations, which we refer to as spin wave precursors.

To illustrate the origin of spin wavepacket precursors, consider the V-shaped dispersion relation for propagation in the $y$ direction shown in Fig. \ref{fig5}(d). In this approximation to the actual relation (Fig. \ref{fig4}(b)), 
SWs propagate with constant velocity for $k_y < 2k_z$, and do not propagate for $k_y>2k_z$. The precursor effects arise from SW modes in which $k_z\sigma$ is small, such that the Gaussian distribution of photoexcited wavevectors (green line in Fig. \ref{fig5}(d)) spans both propagating and non-propagating regimes. 

The time-evolution of the wavepacket is given by summing the contributions from the two regimes,
 \begin{align}
    \delta M(y,t) \propto 2\cos{\omega_0 t} \bigg[&\int_0^{2k_z}dk_y g(k_y) \Big[\cos{k_y(y-vt)}+\cos{k_y(y+vt)}\Big] \nonumber \\
     +&\int^{\infty}_{2k_z}dk_y g(k_y)\cos{(k_y y)} \bigg],
\end{align}
where $\omega_0$ is the frequency at $k_y=0$ and $v$ is the slope of the V-shaped region. Fig. \ref{fig5}(e) shows the propagating and non-propagating terms in Eq. 6 evaluated at $t=0$ (red and blue lines, respectively), together with their sum (green line). The individual terms in Eq. 6 are oscillatory with a slowly decaying envelope, as expected for the Fourier transform of a sharply truncated Gaussian. Notice that the oscillations cancel out under summation, yielding the initial Gaussian wavepacket. However, for $t>0$ the propagating component moves away from the origin at velocity $v$ while the nonpropagating component remains stationary, disrupting the initial cancellation of the two components. This effect manifests as appearance of oscillations in magnetization at large distances within a short time frame. In this simplified picture, a spin wave precursor can be seen at arbitrarily large distances within a time of order of the precession period. In reality, the range of detection will be limited by the rounding of the dispersion neglected in our V-shape approximation; nevertheless precursors will appear on time scales that are not set by the SW velocity. 

\section{Conclusion and Outlook}
We have shown that spin waves in Fe$_3$Sn$_2$ can be optically excited, propagated, and detected across large distances ($> 20\ \mu$m) within short timescales ($< 100$ ps). The arrival of precursors reflects a unique regime of light-matter interaction, resulting from the combination of Gaussian laser excitation and V-shaped magnetostatic spin wave dispersion. The potential for applications points towards ultrafast transmission of spin information across macroscopic distances, not limited by the group velocity of the spin waves. Extending measurements to antiferromagnets should in principle exhibit even stronger precursor effects due to their larger spin wave frequencies. 

\section{Materials and Methods}
\subsection{Crystal growth}
Single crystals of Fe3Sn2 were grown using a Chemical Vapor Transport method with conditions outlined in Ref. \cite{Ye2018}. The resulting crystals tend to be hexagonal thin plates, and optical measurements were performed on as-grown (001) surfaces.

\subsection{Field dependence measurements}
The time-resolved magneto-optic Kerr effect (tr-MOKE) measurements with an out-of-plane magnetic field were performed with 1560 nm pump and 780 nm probe laser pulses generated from a Menlo C-Fiber erbium fiber oscillator operating at a repetition rate of 100 MHz. The pump and probe powers were set to 20 mW and 0.1 mW, and focused onto the sample surface with approximate spot sizes of 20 $\mu$m and 6 $\mu$m, respectively, using an objective lens with a numerical aperture (N.A.) of 0.25. The transient changes in Kerr rotation values were subsequently measured with a balanced photodetection scheme and a lock-in amplifier. The pump laser pulses were modulated at 100 kHz with a photo-elastic modulator (PEM).

\subsection{Propagation measurements}
The non-local propagation experiments were carried out with 514 nm pump and 633 nm probe pulses generated from the ORPHEUS-TWINS optical parametric amplifiers pumped by the Light Conversion CARBIDE Yb-KGW laser amplifier operating at the repetition rate of 600 kHz. Both beams were focused onto the sample sample surface with approximate spot sizes of 6 $\mu$m and 5 $\mu$m, respectively, with incident laser powers fixed at 30 $\mu$W. The position of the pump focus was scanned by adjusting the voltage applied to the 2-axis galvanometer-driven mirrors, which are located at a distance $4f$ ($f=50$ cm) before the entrance aperture of the final objective lens ($\text{N.A.} = 0.25$). A pair of telescope lenses with focal lengths of $f$ are placed equidistant from the galvo mirrors and the objective so that the laser beam steered from the galvo mirrors forms a one-to-one image at the entrance of the objective lens. The pump laser pulses were modulated at 100 kHz with a PEM.

\section{Acknowledgements}
C.L., K.W., J.E.M., and J.O. acknowledge support from the Quantum Materials program under the Director, Office of Science, Office of Basic Energy Sciences, Materials Sciences and Engineering Division, of the U.S. Department of Energy, Contract No. DE-AC02-05CH11231. C.L. and J.O. acknowledge partial support from the Spin Physics program under the Director, Office of Science, Office of Basic Energy Sciences, Materials Sciences and Engineering Division, of the U.S. Department of Energy, Contract No. DE-AC02-76SF00515. Y.S. and J.O. acknowledge support from the Gordon and Betty Moore Foundation’s Emergent Phenomena in Quantum Systems Initiative through Grant GBMF4537 to J.O. at UC Berkeley. Y.-M. Lu acknowledges support from NSF under grant number DMR-2011876.  This work was funded, in part, by the Gordon and Betty Moore Foundation EPiQS Initiative, through Grants GBMF3848 and GBMF9070 to J.G.C. (material synthesis) and NSF grant DMR-2104964 (material analysis). L.Y. acknowledges support by the Tsinghua Education Foundation and STC Center for Integrated Quantum Materials, NSF grant number DMR-1231319.

\section{Author Contributions}
C.L. and J.O. designed research, C.L. performed research and analyzed data, Y.S., K.W., Y.-M.L., J.E.M., and J.O. provided theoretical modeling and analysis, C.L., Y.S., and S.R. performed simulations, L.Y. and J.C. synthesized and characterized the samples, C.L. and J.O. wrote the paper with input from all other authors.

\bibliography{bibliography}

\begin{thebibliography}{27}%
\makeatletter
\providecommand \@ifxundefined [1]{%
 \@ifx{#1\undefined}
}%
\providecommand \@ifnum [1]{%
 \ifnum #1\expandafter \@firstoftwo
 \else \expandafter \@secondoftwo
 \fi
}%
\providecommand \@ifx [1]{%
 \ifx #1\expandafter \@firstoftwo
 \else \expandafter \@secondoftwo
 \fi
}%
\providecommand \natexlab [1]{#1}%
\providecommand \enquote  [1]{``#1''}%
\providecommand \bibnamefont  [1]{#1}%
\providecommand \bibfnamefont [1]{#1}%
\providecommand \citenamefont [1]{#1}%
\providecommand \href@noop [0]{\@secondoftwo}%
\providecommand \href [0]{\begingroup \@sanitize@url \@href}%
\providecommand \@href[1]{\@@startlink{#1}\@@href}%
\providecommand \@@href[1]{\endgroup#1\@@endlink}%
\providecommand \@sanitize@url [0]{\catcode `\\12\catcode `\$12\catcode
  `\&12\catcode `\#12\catcode `\^12\catcode `\_12\catcode `\%12\relax}%
\providecommand \@@startlink[1]{}%
\providecommand \@@endlink[0]{}%
\providecommand \url  [0]{\begingroup\@sanitize@url \@url }%
\providecommand \@url [1]{\endgroup\@href {#1}{\urlprefix }}%
\providecommand \urlprefix  [0]{URL }%
\providecommand \Eprint [0]{\href }%
\providecommand \doibase [0]{https://doi.org/}%
\providecommand \selectlanguage [0]{\@gobble}%
\providecommand \bibinfo  [0]{\@secondoftwo}%
\providecommand \bibfield  [0]{\@secondoftwo}%
\providecommand \translation [1]{[#1]}%
\providecommand \BibitemOpen [0]{}%
\providecommand \bibitemStop [0]{}%
\providecommand \bibitemNoStop [0]{.\EOS\space}%
\providecommand \EOS [0]{\spacefactor3000\relax}%
\providecommand \BibitemShut  [1]{\csname bibitem#1\endcsname}%
\let\auto@bib@innerbib\@empty
\bibitem [{\citenamefont {Kajiwara}\ \emph {et~al.}(2010)\citenamefont
  {Kajiwara}, \citenamefont {Harii}, \citenamefont {Takahashi}, \citenamefont
  {Ohe}, \citenamefont {Uchida}, \citenamefont {Mizuguchi}, \citenamefont
  {Umezawa}, \citenamefont {Kawai}, \citenamefont {Ando}, \citenamefont
  {Takanashi} \emph {et~al.}}]{Kajiwara2010}%
  \BibitemOpen
  \bibfield  {author} {\bibinfo {author} {\bibfnamefont {Y.}~\bibnamefont
  {Kajiwara}}, \bibinfo {author} {\bibfnamefont {K.}~\bibnamefont {Harii}},
  \bibinfo {author} {\bibfnamefont {S.}~\bibnamefont {Takahashi}}, \bibinfo
  {author} {\bibfnamefont {J.-i.}\ \bibnamefont {Ohe}}, \bibinfo {author}
  {\bibfnamefont {K.}~\bibnamefont {Uchida}}, \bibinfo {author} {\bibfnamefont
  {M.}~\bibnamefont {Mizuguchi}}, \bibinfo {author} {\bibfnamefont
  {H.}~\bibnamefont {Umezawa}}, \bibinfo {author} {\bibfnamefont
  {H.}~\bibnamefont {Kawai}}, \bibinfo {author} {\bibfnamefont
  {K.}~\bibnamefont {Ando}}, \bibinfo {author} {\bibfnamefont {K.}~\bibnamefont
  {Takanashi}}, \emph {et~al.},\ }\href@noop {} {\bibfield  {journal} {\bibinfo
   {journal} {Nature}\ }\textbf {\bibinfo {volume} {464}},\ \bibinfo {pages}
  {262} (\bibinfo {year} {2010})}\BibitemShut {NoStop}%
\bibitem [{\citenamefont {Cornelissen}\ \emph {et~al.}(2015)\citenamefont
  {Cornelissen}, \citenamefont {Liu}, \citenamefont {Duine}, \citenamefont
  {Youssef},\ and\ \citenamefont {van Wees}}]{Cornelissen2015}%
  \BibitemOpen
  \bibfield  {author} {\bibinfo {author} {\bibfnamefont {L.~J.}\ \bibnamefont
  {Cornelissen}}, \bibinfo {author} {\bibfnamefont {J.}~\bibnamefont {Liu}},
  \bibinfo {author} {\bibfnamefont {R.~A.}\ \bibnamefont {Duine}}, \bibinfo
  {author} {\bibfnamefont {J.~B.}\ \bibnamefont {Youssef}},\ and\ \bibinfo
  {author} {\bibfnamefont {B.~J.}\ \bibnamefont {van Wees}},\ }\href
  {https://doi.org/10.1038/nphys3465} {\bibfield  {journal} {\bibinfo
  {journal} {Nature Physics}\ }\textbf {\bibinfo {volume} {11}},\ \bibinfo
  {pages} {1022} (\bibinfo {year} {2015})}\BibitemShut {NoStop}%
\bibitem [{\citenamefont {Lebrun}\ \emph {et~al.}(2018)\citenamefont {Lebrun},
  \citenamefont {Ross}, \citenamefont {Bender}, \citenamefont {Qaiumzadeh},
  \citenamefont {Baldrati}, \citenamefont {Cramer}, \citenamefont {Brataas},
  \citenamefont {Duine},\ and\ \citenamefont {Kl{\"a}ui}}]{Lebrun2018}%
  \BibitemOpen
  \bibfield  {author} {\bibinfo {author} {\bibfnamefont {R.}~\bibnamefont
  {Lebrun}}, \bibinfo {author} {\bibfnamefont {A.}~\bibnamefont {Ross}},
  \bibinfo {author} {\bibfnamefont {S.~A.}\ \bibnamefont {Bender}}, \bibinfo
  {author} {\bibfnamefont {A.}~\bibnamefont {Qaiumzadeh}}, \bibinfo {author}
  {\bibfnamefont {L.}~\bibnamefont {Baldrati}}, \bibinfo {author}
  {\bibfnamefont {J.}~\bibnamefont {Cramer}}, \bibinfo {author} {\bibfnamefont
  {A.}~\bibnamefont {Brataas}}, \bibinfo {author} {\bibfnamefont {R.~A.}\
  \bibnamefont {Duine}},\ and\ \bibinfo {author} {\bibfnamefont
  {M.}~\bibnamefont {Kl{\"a}ui}},\ }\href
  {https://doi.org/10.1038/s41586-018-0490-7} {\bibfield  {journal} {\bibinfo
  {journal} {Nature}\ }\textbf {\bibinfo {volume} {561}},\ \bibinfo {pages}
  {222} (\bibinfo {year} {2018})}\BibitemShut {NoStop}%
\bibitem [{\citenamefont {Chumak}\ \emph {et~al.}(2015)\citenamefont {Chumak},
  \citenamefont {Vasyuchka}, \citenamefont {Serga},\ and\ \citenamefont
  {Hillebrands}}]{Chumak2015}%
  \BibitemOpen
  \bibfield  {author} {\bibinfo {author} {\bibfnamefont {A.~V.}\ \bibnamefont
  {Chumak}}, \bibinfo {author} {\bibfnamefont {V.~I.}\ \bibnamefont
  {Vasyuchka}}, \bibinfo {author} {\bibfnamefont {A.~A.}\ \bibnamefont
  {Serga}},\ and\ \bibinfo {author} {\bibfnamefont {B.}~\bibnamefont
  {Hillebrands}},\ }\href {https://doi.org/10.1038/nphys3347} {\bibfield
  {journal} {\bibinfo  {journal} {Nature Physics}\ }\textbf {\bibinfo {volume}
  {11}},\ \bibinfo {pages} {453} (\bibinfo {year} {2015})}\BibitemShut
  {NoStop}%
\bibitem [{\citenamefont {Pirro}\ \emph {et~al.}(2021)\citenamefont {Pirro},
  \citenamefont {Vasyuchka}, \citenamefont {Serga},\ and\ \citenamefont
  {Hillebrands}}]{Pirro2021}%
  \BibitemOpen
  \bibfield  {author} {\bibinfo {author} {\bibfnamefont {P.}~\bibnamefont
  {Pirro}}, \bibinfo {author} {\bibfnamefont {V.~I.}\ \bibnamefont
  {Vasyuchka}}, \bibinfo {author} {\bibfnamefont {A.~A.}\ \bibnamefont
  {Serga}},\ and\ \bibinfo {author} {\bibfnamefont {B.}~\bibnamefont
  {Hillebrands}},\ }\href {https://doi.org/10.1038/s41578-021-00332-w}
  {\bibfield  {journal} {\bibinfo  {journal} {Nature Reviews Materials}\
  }\textbf {\bibinfo {volume} {6}},\ \bibinfo {pages} {1114} (\bibinfo {year}
  {2021})}\BibitemShut {NoStop}%
\bibitem [{\citenamefont {Huang}\ \emph {et~al.}(2017)\citenamefont {Huang},
  \citenamefont {Clark}, \citenamefont {Navarro-Moratalla}, \citenamefont
  {Klein}, \citenamefont {Cheng}, \citenamefont {Seyler}, \citenamefont
  {Zhong}, \citenamefont {Schmidgall}, \citenamefont {McGuire}, \citenamefont
  {Cobden} \emph {et~al.}}]{Huang2017}%
  \BibitemOpen
  \bibfield  {author} {\bibinfo {author} {\bibfnamefont {B.}~\bibnamefont
  {Huang}}, \bibinfo {author} {\bibfnamefont {G.}~\bibnamefont {Clark}},
  \bibinfo {author} {\bibfnamefont {E.}~\bibnamefont {Navarro-Moratalla}},
  \bibinfo {author} {\bibfnamefont {D.~R.}\ \bibnamefont {Klein}}, \bibinfo
  {author} {\bibfnamefont {R.}~\bibnamefont {Cheng}}, \bibinfo {author}
  {\bibfnamefont {K.~L.}\ \bibnamefont {Seyler}}, \bibinfo {author}
  {\bibfnamefont {D.}~\bibnamefont {Zhong}}, \bibinfo {author} {\bibfnamefont
  {E.}~\bibnamefont {Schmidgall}}, \bibinfo {author} {\bibfnamefont {M.~A.}\
  \bibnamefont {McGuire}}, \bibinfo {author} {\bibfnamefont {D.~H.}\
  \bibnamefont {Cobden}}, \emph {et~al.},\ }\href@noop {} {\bibfield  {journal}
  {\bibinfo  {journal} {Nature}\ }\textbf {\bibinfo {volume} {546}},\ \bibinfo
  {pages} {270} (\bibinfo {year} {2017})}\BibitemShut {NoStop}%
\bibitem [{\citenamefont {Lee}\ \emph {et~al.}(2016)\citenamefont {Lee},
  \citenamefont {Lee}, \citenamefont {Ryoo}, \citenamefont {Kang},
  \citenamefont {Kim}, \citenamefont {Kim}, \citenamefont {Park}, \citenamefont
  {Park},\ and\ \citenamefont {Cheong}}]{Lee2016}%
  \BibitemOpen
  \bibfield  {author} {\bibinfo {author} {\bibfnamefont {J.-U.}\ \bibnamefont
  {Lee}}, \bibinfo {author} {\bibfnamefont {S.}~\bibnamefont {Lee}}, \bibinfo
  {author} {\bibfnamefont {J.~H.}\ \bibnamefont {Ryoo}}, \bibinfo {author}
  {\bibfnamefont {S.}~\bibnamefont {Kang}}, \bibinfo {author} {\bibfnamefont
  {T.~Y.}\ \bibnamefont {Kim}}, \bibinfo {author} {\bibfnamefont
  {P.}~\bibnamefont {Kim}}, \bibinfo {author} {\bibfnamefont {C.-H.}\
  \bibnamefont {Park}}, \bibinfo {author} {\bibfnamefont {J.-G.}\ \bibnamefont
  {Park}},\ and\ \bibinfo {author} {\bibfnamefont {H.}~\bibnamefont {Cheong}},\
  }\href@noop {} {\bibfield  {journal} {\bibinfo  {journal} {Nano letters}\
  }\textbf {\bibinfo {volume} {16}},\ \bibinfo {pages} {7433} (\bibinfo {year}
  {2016})}\BibitemShut {NoStop}%
\bibitem [{\citenamefont {Song}\ \emph {et~al.}(2022)\citenamefont {Song},
  \citenamefont {Occhialini}, \citenamefont {Erge{\c{c}}en}, \citenamefont
  {Ilyas}, \citenamefont {Amoroso}, \citenamefont {Barone}, \citenamefont
  {Kapeghian}, \citenamefont {Watanabe}, \citenamefont {Taniguchi},
  \citenamefont {Botana} \emph {et~al.}}]{Song2022}%
  \BibitemOpen
  \bibfield  {author} {\bibinfo {author} {\bibfnamefont {Q.}~\bibnamefont
  {Song}}, \bibinfo {author} {\bibfnamefont {C.~A.}\ \bibnamefont
  {Occhialini}}, \bibinfo {author} {\bibfnamefont {E.}~\bibnamefont
  {Erge{\c{c}}en}}, \bibinfo {author} {\bibfnamefont {B.}~\bibnamefont
  {Ilyas}}, \bibinfo {author} {\bibfnamefont {D.}~\bibnamefont {Amoroso}},
  \bibinfo {author} {\bibfnamefont {P.}~\bibnamefont {Barone}}, \bibinfo
  {author} {\bibfnamefont {J.}~\bibnamefont {Kapeghian}}, \bibinfo {author}
  {\bibfnamefont {K.}~\bibnamefont {Watanabe}}, \bibinfo {author}
  {\bibfnamefont {T.}~\bibnamefont {Taniguchi}}, \bibinfo {author}
  {\bibfnamefont {A.~S.}\ \bibnamefont {Botana}}, \emph {et~al.},\ }\href@noop
  {} {\bibfield  {journal} {\bibinfo  {journal} {Nature}\ }\textbf {\bibinfo
  {volume} {602}},\ \bibinfo {pages} {601} (\bibinfo {year}
  {2022})}\BibitemShut {NoStop}%
\bibitem [{\citenamefont {Sonin}(2010)}]{Sonin2010}%
  \BibitemOpen
  \bibfield  {author} {\bibinfo {author} {\bibfnamefont {E.}~\bibnamefont
  {Sonin}},\ }\href@noop {} {\bibfield  {journal} {\bibinfo  {journal}
  {Advances in Physics}\ }\textbf {\bibinfo {volume} {59}},\ \bibinfo {pages}
  {181} (\bibinfo {year} {2010})}\BibitemShut {NoStop}%
\bibitem [{\citenamefont {Shen}(2021)}]{Shen2021}%
  \BibitemOpen
  \bibfield  {author} {\bibinfo {author} {\bibfnamefont {K.}~\bibnamefont
  {Shen}},\ }\href@noop {} {\bibfield  {journal} {\bibinfo  {journal} {Journal
  of Applied Physics}\ }\textbf {\bibinfo {volume} {129}},\ \bibinfo {pages}
  {223906} (\bibinfo {year} {2021})}\BibitemShut {NoStop}%
\bibitem [{\citenamefont {Qaiumzadeh}\ \emph {et~al.}(2017)\citenamefont
  {Qaiumzadeh}, \citenamefont {Skarsv{\aa}g}, \citenamefont {Holmqvist},\ and\
  \citenamefont {Brataas}}]{Qaiumzadeh2017}%
  \BibitemOpen
  \bibfield  {author} {\bibinfo {author} {\bibfnamefont {A.}~\bibnamefont
  {Qaiumzadeh}}, \bibinfo {author} {\bibfnamefont {H.}~\bibnamefont
  {Skarsv{\aa}g}}, \bibinfo {author} {\bibfnamefont {C.}~\bibnamefont
  {Holmqvist}},\ and\ \bibinfo {author} {\bibfnamefont {A.}~\bibnamefont
  {Brataas}},\ }\href@noop {} {\bibfield  {journal} {\bibinfo  {journal}
  {Physical review letters}\ }\textbf {\bibinfo {volume} {118}},\ \bibinfo
  {pages} {137201} (\bibinfo {year} {2017})}\BibitemShut {NoStop}%
\bibitem [{\citenamefont {Le~Ca{\"e}r}\ \emph {et~al.}(1978)\citenamefont
  {Le~Ca{\"e}r}, \citenamefont {Malaman},\ and\ \citenamefont
  {Roques}}]{LeCaer1978}%
  \BibitemOpen
  \bibfield  {author} {\bibinfo {author} {\bibfnamefont {G.}~\bibnamefont
  {Le~Ca{\"e}r}}, \bibinfo {author} {\bibfnamefont {B.}~\bibnamefont
  {Malaman}},\ and\ \bibinfo {author} {\bibfnamefont {B.}~\bibnamefont
  {Roques}},\ }\href@noop {} {\bibfield  {journal} {\bibinfo  {journal}
  {Journal of Physics F: Metal Physics}\ }\textbf {\bibinfo {volume} {8}},\
  \bibinfo {pages} {323} (\bibinfo {year} {1978})}\BibitemShut {NoStop}%
\bibitem [{\citenamefont {Ye}\ \emph {et~al.}(2018)\citenamefont {Ye},
  \citenamefont {Kang}, \citenamefont {Liu}, \citenamefont {Von~Cube},
  \citenamefont {Wicker}, \citenamefont {Suzuki}, \citenamefont {Jozwiak},
  \citenamefont {Bostwick}, \citenamefont {Rotenberg}, \citenamefont {Bell}
  \emph {et~al.}}]{Ye2018}%
  \BibitemOpen
  \bibfield  {author} {\bibinfo {author} {\bibfnamefont {L.}~\bibnamefont
  {Ye}}, \bibinfo {author} {\bibfnamefont {M.}~\bibnamefont {Kang}}, \bibinfo
  {author} {\bibfnamefont {J.}~\bibnamefont {Liu}}, \bibinfo {author}
  {\bibfnamefont {F.}~\bibnamefont {Von~Cube}}, \bibinfo {author}
  {\bibfnamefont {C.~R.}\ \bibnamefont {Wicker}}, \bibinfo {author}
  {\bibfnamefont {T.}~\bibnamefont {Suzuki}}, \bibinfo {author} {\bibfnamefont
  {C.}~\bibnamefont {Jozwiak}}, \bibinfo {author} {\bibfnamefont
  {A.}~\bibnamefont {Bostwick}}, \bibinfo {author} {\bibfnamefont
  {E.}~\bibnamefont {Rotenberg}}, \bibinfo {author} {\bibfnamefont {D.~C.}\
  \bibnamefont {Bell}}, \emph {et~al.},\ }\href@noop {} {\bibfield  {journal}
  {\bibinfo  {journal} {Nature}\ }\textbf {\bibinfo {volume} {555}},\ \bibinfo
  {pages} {638} (\bibinfo {year} {2018})}\BibitemShut {NoStop}%
\bibitem [{\citenamefont {Kumar}\ \emph {et~al.}(2019)\citenamefont {Kumar},
  \citenamefont {Soh}, \citenamefont {Wang},\ and\ \citenamefont
  {Xiong}}]{Kumar2019}%
  \BibitemOpen
  \bibfield  {author} {\bibinfo {author} {\bibfnamefont {N.}~\bibnamefont
  {Kumar}}, \bibinfo {author} {\bibfnamefont {Y.}~\bibnamefont {Soh}}, \bibinfo
  {author} {\bibfnamefont {Y.}~\bibnamefont {Wang}},\ and\ \bibinfo {author}
  {\bibfnamefont {Y.}~\bibnamefont {Xiong}},\ }\href
  {https://doi.org/10.1103/PhysRevB.100.214420} {\bibfield  {journal} {\bibinfo
   {journal} {Phys. Rev. B}\ }\textbf {\bibinfo {volume} {100}},\ \bibinfo
  {pages} {214420} (\bibinfo {year} {2019})}\BibitemShut {NoStop}%
\bibitem [{\citenamefont {Hiebert}\ \emph {et~al.}(1997)\citenamefont
  {Hiebert}, \citenamefont {Stankiewicz},\ and\ \citenamefont
  {Freeman}}]{Hiebert1997}%
  \BibitemOpen
  \bibfield  {author} {\bibinfo {author} {\bibfnamefont {W.}~\bibnamefont
  {Hiebert}}, \bibinfo {author} {\bibfnamefont {A.}~\bibnamefont
  {Stankiewicz}},\ and\ \bibinfo {author} {\bibfnamefont {M.}~\bibnamefont
  {Freeman}},\ }\href@noop {} {\bibfield  {journal} {\bibinfo  {journal}
  {Physical Review Letters}\ }\textbf {\bibinfo {volume} {79}},\ \bibinfo
  {pages} {1134} (\bibinfo {year} {1997})}\BibitemShut {NoStop}%
\bibitem [{\citenamefont {Acremann}\ \emph {et~al.}(2000)\citenamefont
  {Acremann}, \citenamefont {Back}, \citenamefont {Buess}, \citenamefont
  {Portmann}, \citenamefont {Vaterlaus}, \citenamefont {Pescia},\ and\
  \citenamefont {Melchior}}]{Acremann2000}%
  \BibitemOpen
  \bibfield  {author} {\bibinfo {author} {\bibfnamefont {Y.}~\bibnamefont
  {Acremann}}, \bibinfo {author} {\bibfnamefont {C.~H.}\ \bibnamefont {Back}},
  \bibinfo {author} {\bibfnamefont {M.}~\bibnamefont {Buess}}, \bibinfo
  {author} {\bibfnamefont {O.}~\bibnamefont {Portmann}}, \bibinfo {author}
  {\bibfnamefont {A.}~\bibnamefont {Vaterlaus}}, \bibinfo {author}
  {\bibfnamefont {D.}~\bibnamefont {Pescia}},\ and\ \bibinfo {author}
  {\bibfnamefont {H.}~\bibnamefont {Melchior}},\ }\href@noop {} {\bibfield
  {journal} {\bibinfo  {journal} {Science}\ }\textbf {\bibinfo {volume}
  {290}},\ \bibinfo {pages} {492} (\bibinfo {year} {2000})}\BibitemShut
  {NoStop}%
\bibitem [{\citenamefont {Kimel}\ \emph {et~al.}(2004)\citenamefont {Kimel},
  \citenamefont {Kirilyuk}, \citenamefont {Tsvetkov}, \citenamefont {Pisarev},\
  and\ \citenamefont {Rasing}}]{Kimel2004}%
  \BibitemOpen
  \bibfield  {author} {\bibinfo {author} {\bibfnamefont {A.}~\bibnamefont
  {Kimel}}, \bibinfo {author} {\bibfnamefont {A.}~\bibnamefont {Kirilyuk}},
  \bibinfo {author} {\bibfnamefont {A.}~\bibnamefont {Tsvetkov}}, \bibinfo
  {author} {\bibfnamefont {R.}~\bibnamefont {Pisarev}},\ and\ \bibinfo {author}
  {\bibfnamefont {T.}~\bibnamefont {Rasing}},\ }\href@noop {} {\bibfield
  {journal} {\bibinfo  {journal} {Nature}\ }\textbf {\bibinfo {volume} {429}},\
  \bibinfo {pages} {850} (\bibinfo {year} {2004})}\BibitemShut {NoStop}%
\bibitem [{\citenamefont {Damon}\ and\ \citenamefont
  {Eshbach}(1961)}]{Damon1961}%
  \BibitemOpen
  \bibfield  {author} {\bibinfo {author} {\bibfnamefont {R.}~\bibnamefont
  {Damon}}\ and\ \bibinfo {author} {\bibfnamefont {J.}~\bibnamefont
  {Eshbach}},\ }\href
  {https://doi.org/https://doi.org/10.1016/0022-3697(61)90041-5} {\bibfield
  {journal} {\bibinfo  {journal} {Journal of Physics and Chemistry of Solids}\
  }\textbf {\bibinfo {volume} {19}},\ \bibinfo {pages} {308} (\bibinfo {year}
  {1961})}\BibitemShut {NoStop}%
\bibitem [{\citenamefont {Stancil}\ and\ \citenamefont
  {Prabhakar}(2009)}]{Stancil2009}%
  \BibitemOpen
  \bibfield  {author} {\bibinfo {author} {\bibfnamefont {D.}~\bibnamefont
  {Stancil}}\ and\ \bibinfo {author} {\bibfnamefont {A.}~\bibnamefont
  {Prabhakar}},\ }\href {https://books.google.com/books?id=ehN6-ubvKwoC} {\emph
  {\bibinfo {title} {Spin Waves: Theory and Applications}}}\ (\bibinfo
  {publisher} {Springer US},\ \bibinfo {year} {2009})\BibitemShut {NoStop}%
\bibitem [{\citenamefont {Bae}\ \emph {et~al.}(2022)\citenamefont {Bae},
  \citenamefont {Wang}, \citenamefont {Scheie}, \citenamefont {Xu},
  \citenamefont {Chica}, \citenamefont {Diederich}, \citenamefont {Cenker},
  \citenamefont {Ziebel}, \citenamefont {Bai}, \citenamefont {Ren} \emph
  {et~al.}}]{Bae2022}%
  \BibitemOpen
  \bibfield  {author} {\bibinfo {author} {\bibfnamefont {Y.~J.}\ \bibnamefont
  {Bae}}, \bibinfo {author} {\bibfnamefont {J.}~\bibnamefont {Wang}}, \bibinfo
  {author} {\bibfnamefont {A.}~\bibnamefont {Scheie}}, \bibinfo {author}
  {\bibfnamefont {J.}~\bibnamefont {Xu}}, \bibinfo {author} {\bibfnamefont
  {D.~G.}\ \bibnamefont {Chica}}, \bibinfo {author} {\bibfnamefont {G.~M.}\
  \bibnamefont {Diederich}}, \bibinfo {author} {\bibfnamefont {J.}~\bibnamefont
  {Cenker}}, \bibinfo {author} {\bibfnamefont {M.~E.}\ \bibnamefont {Ziebel}},
  \bibinfo {author} {\bibfnamefont {Y.}~\bibnamefont {Bai}}, \bibinfo {author}
  {\bibfnamefont {H.}~\bibnamefont {Ren}}, \emph {et~al.},\ }\href@noop {}
  {\bibfield  {journal} {\bibinfo  {journal} {Nature}\ }\textbf {\bibinfo
  {volume} {609}},\ \bibinfo {pages} {282} (\bibinfo {year}
  {2022})}\BibitemShut {NoStop}%
\bibitem [{\citenamefont {Langner}\ \emph {et~al.}(2009)\citenamefont
  {Langner}, \citenamefont {Kantner}, \citenamefont {Chu}, \citenamefont
  {Martin}, \citenamefont {Yu}, \citenamefont {Seidel}, \citenamefont
  {Ramesh},\ and\ \citenamefont {Orenstein}}]{Langner2009}%
  \BibitemOpen
  \bibfield  {author} {\bibinfo {author} {\bibfnamefont {M.~C.}\ \bibnamefont
  {Langner}}, \bibinfo {author} {\bibfnamefont {C.~L.~S.}\ \bibnamefont
  {Kantner}}, \bibinfo {author} {\bibfnamefont {Y.~H.}\ \bibnamefont {Chu}},
  \bibinfo {author} {\bibfnamefont {L.~M.}\ \bibnamefont {Martin}}, \bibinfo
  {author} {\bibfnamefont {P.}~\bibnamefont {Yu}}, \bibinfo {author}
  {\bibfnamefont {J.}~\bibnamefont {Seidel}}, \bibinfo {author} {\bibfnamefont
  {R.}~\bibnamefont {Ramesh}},\ and\ \bibinfo {author} {\bibfnamefont
  {J.}~\bibnamefont {Orenstein}},\ }\href
  {https://doi.org/10.1103/PhysRevLett.102.177601} {\bibfield  {journal}
  {\bibinfo  {journal} {Phys. Rev. Lett.}\ }\textbf {\bibinfo {volume} {102}},\
  \bibinfo {pages} {177601} (\bibinfo {year} {2009})}\BibitemShut {NoStop}%
\bibitem [{\citenamefont {Suhl}(1955)}]{Suhl1955}%
  \BibitemOpen
  \bibfield  {author} {\bibinfo {author} {\bibfnamefont {H.}~\bibnamefont
  {Suhl}},\ }\href {https://doi.org/10.1103/PhysRev.97.555.2} {\bibfield
  {journal} {\bibinfo  {journal} {Phys. Rev.}\ }\textbf {\bibinfo {volume}
  {97}},\ \bibinfo {pages} {555} (\bibinfo {year} {1955})}\BibitemShut
  {NoStop}%
\bibitem [{\citenamefont {Demokritov}\ \emph {et~al.}(1989)\citenamefont
  {Demokritov}, \citenamefont {Kreines}, \citenamefont {Kudinov},\ and\
  \citenamefont {Petrov}}]{Demokritov1989}%
  \BibitemOpen
  \bibfield  {author} {\bibinfo {author} {\bibfnamefont {S.~.}\ \bibnamefont
  {Demokritov}}, \bibinfo {author} {\bibfnamefont {N.~M.}\ \bibnamefont
  {Kreines}}, \bibinfo {author} {\bibfnamefont {V.~I.}\ \bibnamefont
  {Kudinov}},\ and\ \bibinfo {author} {\bibfnamefont {S.~V.}\ \bibnamefont
  {Petrov}},\ }\href@noop {} {\bibfield  {journal} {\bibinfo  {journal} {Zh.
  Eksp. Teor. Fiz.}\ }\textbf {\bibinfo {volume} {95}},\ \bibinfo {pages}
  {2211} (\bibinfo {year} {1989})}\BibitemShut {NoStop}%
\bibitem [{\citenamefont {Satoh}\ \emph {et~al.}(2012)\citenamefont {Satoh},
  \citenamefont {Terui}, \citenamefont {Moriya}, \citenamefont {Ivanov},
  \citenamefont {Ando}, \citenamefont {Saitoh}, \citenamefont {Shimura},\ and\
  \citenamefont {Kuroda}}]{Satoh2012}%
  \BibitemOpen
  \bibfield  {author} {\bibinfo {author} {\bibfnamefont {T.}~\bibnamefont
  {Satoh}}, \bibinfo {author} {\bibfnamefont {Y.}~\bibnamefont {Terui}},
  \bibinfo {author} {\bibfnamefont {R.}~\bibnamefont {Moriya}}, \bibinfo
  {author} {\bibfnamefont {B.~A.}\ \bibnamefont {Ivanov}}, \bibinfo {author}
  {\bibfnamefont {K.}~\bibnamefont {Ando}}, \bibinfo {author} {\bibfnamefont
  {E.}~\bibnamefont {Saitoh}}, \bibinfo {author} {\bibfnamefont
  {T.}~\bibnamefont {Shimura}},\ and\ \bibinfo {author} {\bibfnamefont
  {K.}~\bibnamefont {Kuroda}},\ }\href@noop {} {\bibfield  {journal} {\bibinfo
  {journal} {Nature Photonics}\ }\textbf {\bibinfo {volume} {6}},\ \bibinfo
  {pages} {662} (\bibinfo {year} {2012})}\BibitemShut {NoStop}%
\bibitem [{\citenamefont {Dally}\ \emph {et~al.}(2021)\citenamefont {Dally},
  \citenamefont {Phelan}, \citenamefont {Bishop}, \citenamefont {Ghimire},\
  and\ \citenamefont {Lynn}}]{Dally2021}%
  \BibitemOpen
  \bibfield  {author} {\bibinfo {author} {\bibfnamefont {R.~L.}\ \bibnamefont
  {Dally}}, \bibinfo {author} {\bibfnamefont {D.}~\bibnamefont {Phelan}},
  \bibinfo {author} {\bibfnamefont {N.}~\bibnamefont {Bishop}}, \bibinfo
  {author} {\bibfnamefont {N.~J.}\ \bibnamefont {Ghimire}},\ and\ \bibinfo
  {author} {\bibfnamefont {J.~W.}\ \bibnamefont {Lynn}},\ }\href@noop {}
  {\bibfield  {journal} {\bibinfo  {journal} {Crystals}\ }\textbf {\bibinfo
  {volume} {11}},\ \bibinfo {pages} {307} (\bibinfo {year} {2021})}\BibitemShut
  {NoStop}%
\bibitem [{\citenamefont {Hurben}\ and\ \citenamefont
  {Patton}(1996)}]{Hurben1996}%
  \BibitemOpen
  \bibfield  {author} {\bibinfo {author} {\bibfnamefont {M.}~\bibnamefont
  {Hurben}}\ and\ \bibinfo {author} {\bibfnamefont {C.}~\bibnamefont
  {Patton}},\ }\href@noop {} {\bibfield  {journal} {\bibinfo  {journal}
  {Journal of Magnetism and Magnetic Materials}\ }\textbf {\bibinfo {volume}
  {163}},\ \bibinfo {pages} {39} (\bibinfo {year} {1996})}\BibitemShut
  {NoStop}%
\bibitem [{\citenamefont {Hashimoto}\ \emph {et~al.}(2017)\citenamefont
  {Hashimoto}, \citenamefont {Daimon}, \citenamefont {Iguchi}, \citenamefont
  {Oikawa}, \citenamefont {Shen}, \citenamefont {Sato}, \citenamefont
  {Bossini}, \citenamefont {Tabuchi}, \citenamefont {Satoh}, \citenamefont
  {Hillebrands} \emph {et~al.}}]{Hashimoto2017}%
  \BibitemOpen
  \bibfield  {author} {\bibinfo {author} {\bibfnamefont {Y.}~\bibnamefont
  {Hashimoto}}, \bibinfo {author} {\bibfnamefont {S.}~\bibnamefont {Daimon}},
  \bibinfo {author} {\bibfnamefont {R.}~\bibnamefont {Iguchi}}, \bibinfo
  {author} {\bibfnamefont {Y.}~\bibnamefont {Oikawa}}, \bibinfo {author}
  {\bibfnamefont {K.}~\bibnamefont {Shen}}, \bibinfo {author} {\bibfnamefont
  {K.}~\bibnamefont {Sato}}, \bibinfo {author} {\bibfnamefont {D.}~\bibnamefont
  {Bossini}}, \bibinfo {author} {\bibfnamefont {Y.}~\bibnamefont {Tabuchi}},
  \bibinfo {author} {\bibfnamefont {T.}~\bibnamefont {Satoh}}, \bibinfo
  {author} {\bibfnamefont {B.}~\bibnamefont {Hillebrands}}, \emph {et~al.},\
  }\href@noop {} {\bibfield  {journal} {\bibinfo  {journal} {Nature
  communications}\ }\textbf {\bibinfo {volume} {8}},\ \bibinfo {pages} {1}
  (\bibinfo {year} {2017})}\BibitemShut {NoStop}%
\end{thebibliography}%


\begin{thebibliography}{1}%
\makeatletter
\providecommand \@ifxundefined [1]{%
 \@ifx{#1\undefined}
}%
\providecommand \@ifnum [1]{%
 \ifnum #1\expandafter \@firstoftwo
 \else \expandafter \@secondoftwo
 \fi
}%
\providecommand \@ifx [1]{%
 \ifx #1\expandafter \@firstoftwo
 \else \expandafter \@secondoftwo
 \fi
}%
\providecommand \natexlab [1]{#1}%
\providecommand \enquote  [1]{``#1''}%
\providecommand \bibnamefont  [1]{#1}%
\providecommand \bibfnamefont [1]{#1}%
\providecommand \citenamefont [1]{#1}%
\providecommand \href@noop [0]{\@secondoftwo}%
\providecommand \href [0]{\begingroup \@sanitize@url \@href}%
\providecommand \@href[1]{\@@startlink{#1}\@@href}%
\providecommand \@@href[1]{\endgroup#1\@@endlink}%
\providecommand \@sanitize@url [0]{\catcode `\\12\catcode `\$12\catcode
  `\&12\catcode `\#12\catcode `\^12\catcode `\_12\catcode `\%12\relax}%
\providecommand \@@startlink[1]{}%
\providecommand \@@endlink[0]{}%
\providecommand \url  [0]{\begingroup\@sanitize@url \@url }%
\providecommand \@url [1]{\endgroup\@href {#1}{\urlprefix }}%
\providecommand \urlprefix  [0]{URL }%
\providecommand \Eprint [0]{\href }%
\providecommand \doibase [0]{https://doi.org/}%
\providecommand \selectlanguage [0]{\@gobble}%
\providecommand \bibinfo  [0]{\@secondoftwo}%
\providecommand \bibfield  [0]{\@secondoftwo}%
\providecommand \translation [1]{[#1]}%
\providecommand \BibitemOpen [0]{}%
\providecommand \bibitemStop [0]{}%
\providecommand \bibitemNoStop [0]{.\EOS\space}%
\providecommand \EOS [0]{\spacefactor3000\relax}%
\providecommand \BibitemShut  [1]{\csname bibitem#1\endcsname}%
\let\auto@bib@innerbib\@empty
\bibitem [{\citenamefont {Dally}\ \emph {et~al.}(2021)\citenamefont {Dally},
  \citenamefont {Phelan}, \citenamefont {Bishop}, \citenamefont {Ghimire},\
  and\ \citenamefont {Lynn}}]{Dally2021}%
  \BibitemOpen
  \bibfield  {author} {\bibinfo {author} {\bibfnamefont {R.~L.}\ \bibnamefont
  {Dally}}, \bibinfo {author} {\bibfnamefont {D.}~\bibnamefont {Phelan}},
  \bibinfo {author} {\bibfnamefont {N.}~\bibnamefont {Bishop}}, \bibinfo
  {author} {\bibfnamefont {N.~J.}\ \bibnamefont {Ghimire}},\ and\ \bibinfo
  {author} {\bibfnamefont {J.~W.}\ \bibnamefont {Lynn}},\ }\href@noop {}
  {\bibfield  {journal} {\bibinfo  {journal} {Crystals}\ }\textbf {\bibinfo
  {volume} {11}},\ \bibinfo {pages} {307} (\bibinfo {year} {2021})}\BibitemShut
  {NoStop}%
\end{thebibliography}%
\end{document}


\title{Supporting Information for ``Spin wavepackets in the Kagome ferromagnet Fe$_3$Sn$_2$: propagation and precursors"}
\maketitle
\section{Symmetry-based microscopic model}
Fe$_3$Sn$_2$ has space group $R\bar3m$ (\#166), in which the magnetic Fe$^{x+}$ ions form a kagome lattice in each layer that are A-B-C stacked in the three-dimensional crystalline structure. Our goal is to describe the spin dynamics of Fe$_3$Sn$_2$ by the local moments $\{\vec S_r\}$ of Fe$^{x+}$ ions (on lattice site $r$). Although there are also itinerant electrons present, their contribution to the interactions between Fe$^{x+}$ moments can be incorporated by integrating them out to obtain a pure spin model of local moments $\{\vec S_r\}$. 

Since Fe$_3$Sn$_2$ is in a ferromagnetic ordered phase below $T_c=657$~K, we expect a ferromagnetic coupling between kagome layers, and therefore focus our analysis on a monolayer model on the two-dimensional kagome lattice. The minimal model compatible with the space group symmetry is the XXZ Hamiltonian:
\bea
\hat H_{XXZ}=-J\sum_{\expval{i,j}}S_i^xS_j^x+S_i^yS_j^y+\Delta S_i^zS_j^z,
\eea
where $\Delta$ is a parameter characterizing the easy-plane anisotropy of the compound. Specifically, when $\Delta<1$, the ferromagnet favors an in-plane magnetic moment, while $\Delta>1$ describes an easy-axis ferromagnet whose magnetic moments align along the c-axis. We have neglected the ``breathing'' terms that distinguish the upper and lower triangle of the kagome lattice, as it is not necessary to explain the observed magnon properties. 

Given the spin model, one can use a linear spin wave (LSW) approach to study the magnon dispersion of the system. In the XXZ model, if the ground state is an easy-plane ferromagnet, the magnetic moments can point to any direction in the $a-b$ plane, giving rise to gapless Goldstone modes of the spontaneously broken continuous symmetry $U(1)=\{e^{\imth\theta\sum_rS^z_r}|0\leq\theta<2\pi\}$. In fact, in the framework of LSW approach, this gapless $\vec k=0$ magnon mode can persist even without a continuous $U(1)$ symmetry of the spin model, summarized in the following statement:

{\bf Theorem:} \emph{In a bilinear spin model of a spin-orbit coupled magnet, if the long range ferromagnetic order spontaneously breaks the $n$-fold crystalline rotation symmetry $C_n$ with $n=3,4,6$, there will be gapless modes in the magnon spectrum near the zone center $\vec k=0$ using the LSW approach.}

Below we explain the reason behind this theorem. Consider a blinear spin model with $C_n$ rotational symmetry along the c-axis:
\bea
\hat H_{2}=\sum_{i,j}\sum_{a,b,=x,y,z}S_i^aR_{i,j}^{a,b}S_j^b,~~~S_{\vec r}^+=S_{\vec r}^x+\imth S_{\vec r}^y\overset{C_n}\longrightarrow e^{2\pi\imth/n}S^+_{C_n\vec r}.
\eea
In the ferromagnetic order with ordering wavevector $\vec Q=0$, the spin raising operator $S^+$ near the zone center $\Gamma$ transforms as, 
\bea
S^+_{\vec k=0}=\frac1{\sqrt N}\sum_rS^+_r\overset{C_n}\longrightarrow e^{2\pi\imth/n}S^+_{\vec k=0}.
\eea
Therefore near the zone center $\vec k\approx0$, any $C_n$-symmetric bilinear Hamiltonian must have the following form:
\bea
\hat H_2=\sum_{|\vec k|\ll1}f({\vec k})S^+_{\vec k}S^-_{\vec k}+h.c.+\cdots
\eea
This means that in a generic bilinear spin model, the $C_n$ rotational symmetry for $n=3,4,6$ is enlarged to an emergent $U(1)$ symmetry in the long wavelength ($|\vec k|\ll1$) limit. As a result, spontaneously breaking the $C_n$ symmetry in an in-plane ferromagnetic order will also break the emergent $U(1)$ symmetry near zone center, hence giving rise to gapless magnon modes. These gapless magnons can be thought of as the ``pseudo-Goldstone'' modes from spontaneously breaking the emergent $U(1)$ symmetry. 

The gapless magnon modes, however, are inconsistent with the gap of 8 GHz $\approx0.2$ meV that we observe in Fe$_3$Sn$_2$. As a result, within the LSW theory, we have to go beyond the bilinear XXZ model described above. Due to Hermiticity and time reversal symmetry, any term in the Hamiltonian must contain an even number of spin Hamiltonians. To reduce the emergent $U(1)$ symmetry down to $C_n$, we have to introduce four-spin or six-spin interactions. The lowest order term compatible with the space group $R\bar3m$ is a quartic term
\bea
\hat H_4(\vec k=0)\sim f_4(\vec k=0)(S_{\vec k=0}^+)^3S^z_{\vec k=0}+h.c.,
\eea
Unfortunately, for in-plane magnetic order, such a term vanishes in the LSW approximation and doesn't open up a magnon gap at the zone center. As a result, the lowest-order term to explain the magnon gap at $\Gamma$ is a six-spin term $\sim(S^+)^6+h.c.$. In Fe$_3$Sn$_2$, the simplest realization of this term is a 3-site ring exchange interaction:
\bea
\hat H_{ring}=-K\sum_{\expval{i,j,k}}\big[(S_i^+)^2(S_j^+)^2(S_k^+)^2+h.c.\big],
\eea
where $\expval{i,j,k}$ denotes nearest neighbor triplets $i,j,k$ on the same triangle. 

To summarize, our minimal model to describe the spin dynamics of Fe$_3$Sn$_2$ is given as follows:
\bea\notag
&\hat H_{min}=\hat H_{XXZ}+\hat H_{ring}+\hat H_{Zeeman}\\
&=-J\sum_{\expval{i,j}}(S_i^xS_j^x+S_i^yS_j^y+\Delta S_i^zS_j^z)-K\sum_{\expval{i,j,k}}\big[(S_i^+)^2(S_j^+)^2(S_k^+)^2+h.c.\big]-B_z\sum_iS^z_i,
\eea
where $B_z$ labels the Zeeman field applied along $c-$axis. 
At $\vec k=0$, with $B_z=0$, we can work out the magnon spectrum using a Holstein-Primakoff transformation:
\begin{align}
S^z_i&=S-b^\dagger_i b_i\\
S^+_i&=\sqrt{2S-b^\dagger_i b_i}\cdot b_i\approx\sqrt{2S}\cdot b_i\\
S^-_i&=b^\dagger_i\sqrt{2S-b^\dagger_i b_i}\approx b^\dagger_i\sqrt{2S}
\end{align}
where $b_i^\dagger,b_i$ are boson creation and annihilation operators, and $S$ is the total spin on each site $i$. After making these substitutions, isolating quadratic terms, and converting to momentum space, we can write the boson BdG Hamiltonian as
\begin{align}
    \hat H_{LSW}=\sum_{\vec k} \hat\Gamma^\dagger(\vec k)\begin{pmatrix}
        F(\vec k) & G(\vec k)\\
        G^*(-\vec k) & F^*(-\vec k)
    \end{pmatrix}
    \hat\Gamma(\vec k)
    \equiv\sum_{\vec k} \hat\Gamma^\dagger(\vec k)H_b(\vec k)
    \hat\Gamma(\vec k),
\end{align}
where
\begin{align}
    \hat\Gamma^\dagger(\vec k)=\begin{pmatrix}
        \hat b^\dagger_{\vec k 1}&\ldots&\hat b^\dagger_{\vec k m}&\hat b_{-\vec k 1}&\ldots&\hat b_{ -\vec k m}
    \end{pmatrix}
\end{align}
and there are $m=3$ spins in each unit cell. At $\vec k =0$, we have the simple form
\begin{align}
    F(\vec k=0)=\begin{pmatrix}
        A & B & B \\
        B & A & B \\
        B & B & A \\
    \end{pmatrix}\\
    G(\vec k=0)=\begin{pmatrix}
        C & D & D \\
        D & C & D \\
        D & D & C \\
    \end{pmatrix},
\end{align}
where
\begin{align}
    A&=2JS+\frac{3}{2}KS^3(2S-1)^2\\
    B&=-JS(1+\frac{\delta}{2})+2KS^4(2S-1)\\
    C&=-\frac{1}{2}KS^3(2S-1)^2\\
    D&=-\frac{1}{2}JS\delta-2KS^4(2S-1),
\end{align}
where we define $\delta\equiv \Delta-1$. To find the magnon spectrum, we diagonalize the matrix $\sigma_z H_b$, where $\sigma_z$ acts on the $(\hat b, \hat b^\dagger)$ space. We find that two of the bands are degenerate at $\vec k=0$
\begin{align}
    E_1&=\sqrt{(A+2B)^2-(C+2D)^2}\\
    E_{2,3}&=\sqrt{(A-B)^2-(C-D)^2}.
\end{align}
For small values $\frac{K'}{J},|\delta|\ll1$, the non-degenerate band is lower in energy, with energy given by
\begin{align}
    E_1=\sqrt{2K'(6S-1)(-2JS\delta+(2S-1)K')}=2S^2\sqrt{KJ(2S-1)(6S-1)}\Big[1-\delta+\frac{S^2(2S-1)^2K}{2J}\Big]^{1/2}
\end{align}
where $K'\equiv KS^3(2S-1)$. The experimentally determined spin stiffness of Fe$_3$Sn$_2$ is \cite{Dally2021}
\bea
D = 2JSa^2 \approx 231 \text{ meV \AA$^{2}$},
\eea
where $a=5.34 \text{ \AA}$ is the the lattice constant. Given the gap $E_1\approx8$ GHz $\approx33~\mu$eV, one can estimate the exchange coupling to be
\bea
K\sim\frac{E_1^2}{JS^6}\sim\frac{E_1^2}{DS^5/2a^2}\sim0.27~\mu \text{eV}
\eea
where we have set $S\sim O(1)$ in the estimation. This means a very small and realistic ring exchange coupling can already induce the $0.03$~meV magnon gap at the zone center.  

\section{Free energy analysis}

Consider a classical ferromagnetic ground state with 
\bea
\expval{\vec S_r}=(M_x,M_y,M_z),~~~\forall~r\in\Lambda
\eea
the free energy density is given by
\bea
\mathcal{F}=\frac{\expval{\hat H_{min}}}{N}=-6J(M_x^2+M_y^2+\Delta M_z^2)-2K\big[(M_x+\imth M_y)^6+h.c.\big]-3B_zM_z,
\eea
where $N$ is the total number of unit cells in the system. Making use the relation, 
\bea\label{total spin}
(\vec S_r)^2=S(S+1)=M_x^2+M_y^2+M_z^2,
\eea
we can expand around the in-plane easy-axis ferromagnetic order $\expval{\vec S_r}=\sqrt{S(S+1)}(1,0,0)$ and do the LSW expansion,
\bea
M_x=\sqrt{S(S+1)-M_y^2-M_z^2}\approx\sqrt{S(S+1)}-\frac{M_y^2+M_z^2}{2\sqrt{S(S+1)}}+O(\frac{M_{y,z}}{S})^4,
\eea
This yields the zero-field free energy density as a function of spin wave fluctuations $M_{y,z}\ll1$:
\bea
&\mathcal{F}_{LSW}(B_z=0)=\rho_yM_y^2+\rho_zM_z^2+O(M^4_{y,z}),\\
&\rho_y=72[S(S+1)]^2K,~~~\rho_z=6J(1-\Delta)+12[S(S+1)]^2K,
\eea
where $S$ is the total spin of each Fe$^{x+}$ ion. Making use of relation (\ref{total spin}) we can rewrite the above free energy density as
\bea
\mathcal{F}_{LSW}(B_z=0)=-\rho_y M_x^2+(\rho_z-\rho_y)M_z^2+O(M^2_{y,z}).
\eea

\section{Angular dependence of the spin wave propagation}
\begin{figure} [ht!]
\includegraphics[width=6.5in]{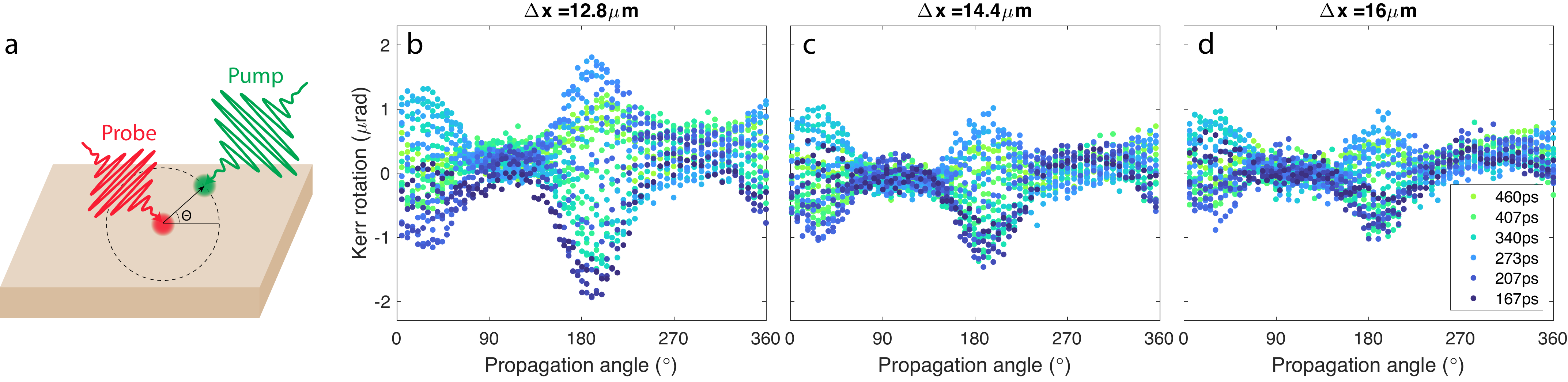}
\caption{\label{figS1} Angular dependence of the spin wave propagation. (a) Overview of the experimental setup where the orientation angle between the pump and probe beams is being continuously varied. (b)-(d) Angle-dependent tr-MOKE measurements taken at pump-probe separation distances of 12.8, 14.4, and 16 $\mu$m, respectively. Different colors indicate various time delays between the pump and probe pulses.}
\end{figure}

In this experiment, the transient changes in Kerr rotation were measured as a function of angular orientation between the pump and probe beams, while the spatial separation and time delay between the two pulses are fixed (Fig. \ref{figS1}(a)). The data acquired at various separation distances and time delays between the pump and probe pulses (Fig. \ref{figS1}(b)-(d)) show that the major propagation directions of the spin waves are along $\sim 20^{\circ}$ and $200^{\circ}$ with respect to the $x$ axis, in agreement with the bi-directional propagation of forward volume modes.

\section{Derivation of the dipolar magnon dispersion}
\label{zero_field}
The magneto-crystalline anisotropy can be simplified to the combination of a hard axis along the out-of-plane direction ($z$-axis) and an easy axis along the in-plane direction ($x$-axis). The magnetization vector is denoted as $\bm{M}=\left(M_{x}, M_{y}, M_{z}\right)$. The free energy of the anisotropy $F_{A}$ can be written as
\begin{equation}
F_{A}=(-K_{x}M_{x}^{2}+K_{z}M_{z}^{2})/M_{s}^{2},
\end{equation}
where $K_{x}$ is the in-plane anisotropy energy, $K_{z}$ is the out-of-plane anisotropy energy ($K_{x}$, $K_{z}>0$) and $M_{s}$ is the saturation magnetization. The effective field induced by the anisotropy energy can be derived as
\begin{equation}
\bm{H}_{A}=-\nabla_{\bm{M}}F_{A}=-\frac{2}{M_{s}^{2}}(K_{x}M_{x}\hat{x}-K_{z}M_{z}\hat{z}).
\end{equation}
In addition to $\bm{H}_A$, the effective field $\bm{H}_{eff}$ includes the dynamic demagnetizing field $\bm{h}$, which is the key to consider the magnetic dipole-dipole interaction. Therefore, our goal is to solve the Landau-Lifshitz equation below to calculate the magnon dispersion,
\begin{equation}
\label{eq:LL}
\begin{aligned}
&\frac{d\bm{M}}{dt}=-\gamma\bm{M}\times\bm{H}_\text{eff},
\\& \bm{H}_\text{eff} = \bm{H}_{A} + \bm{h},
\end{aligned}
\end{equation}
where $\gamma$ is the gyromagnetic ratio. We assume that the equilibrium magnetization is along the $x$-direction and consider an infinitely large sample, so that the normal modes have the form of plane waves. We have
\begin{equation}
\label{eq:planewave}
\begin{aligned}
&M_{x}=M_{s},
\\&M_{y} = m_{y}(\bm{r},t) = m_{y}e^{i(\bm{k}\cdot\bm{r}-\omega t)},
\\&M_{z} = m_{z}(\bm{r},t) = m_{z}e^{i(\bm{k}\cdot\bm{r}-\omega t)},
\\&h_{x}(\bm{r},t) = h_{x}e^{i(\bm{k}\cdot\bm{r}-\omega t)},
\\&h_{y}(\bm{r},t) = h_{y}e^{i(\bm{k}\cdot\bm{r}-\omega t)},
\\&h_{z}(\bm{r},t) = h_{z}e^{i(\bm{k}\cdot\bm{r}-\omega t)}.
\end{aligned}
\end{equation}
Since $m_{y,z}$ and $h_{x,y,z}$ are the dynamic magnetization and dynamic magnetic fields, $m_{y,z}$, $h_{x,y,z}<<M_{s}$. Substituting Eq.~\ref{eq:planewave} into Eq.~\ref{eq:LL}, we get the relation between $m_{y,z}$ and $h_{y,z}$.
\begin{equation}
\label{eq:m-h-relation}
\begin{bmatrix}
m_{y} \\ m_{z}
\end{bmatrix}=\frac{\gamma M_{s}}{\omega_{T}\omega_{x}-\omega^{2}}
\begin{bmatrix}
\omega_{T} & -i\omega \\
i\omega & \omega_{x}
\end{bmatrix}
\begin{bmatrix}
h_{y} \\ h_{z}
\end{bmatrix},
\end{equation}
where
\begin{equation}
\omega_{x} = \frac{2\gamma K_{x}}{M_{s}}; \ \omega_{T} = \frac{2\gamma K_{T}}{M_{s}};\ K_{T} = K_{x} + K_{z}.
\end{equation}
Noting the relation $\nabla\times \bm{h} = 0$, we can consider the dynamic magnetic field as the gradient of a scalar potential $\psi$,
\begin{equation}
\begin{aligned}
&\psi = a e^{i(\bm{k}\cdot\bm{r}-\omega t)},
\\& \bm{h} = \nabla \psi = i\bm{k}a e^{i(\bm{k}\cdot\bm{r}-\omega t)}.
\end{aligned}
\end{equation}
Taking the dipole-dipole interaction into account, we use the Gauss's Law of the Maxwell's equations
\begin{equation}
\label{eq:Maxwell}
\nabla\cdot \bm{b}=0,
\end{equation}
where $\bm{b} = \bm{h}+4\pi\bm{M}$. Using the relation in Eq.~\ref{eq:m-h-relation}, we have
\begin{equation}
\begin{aligned}
&b_{x} = h_{x}+4\pi M_{s},
\\&b_{y} = (1+\frac{\omega_{M}\omega_{T}}{\omega_{T}\omega_{x}-\omega^{2}})h_{y}-\frac{i\omega_{M}\omega}{\omega_{T}\omega_{x}-\omega^{2}}h_{z},
\\&b_{z} = \frac{i\omega_{M}\omega}{\omega_{T}\omega_{x}-\omega^{2}}h_{y}+(1+\frac{\omega_{M}\omega_{x}}{\omega_{T}\omega_{x}-\omega^{2}})h_{z},
\end{aligned}
\end{equation}
where $\omega_{M}=4\pi\gamma M_{s}$.
Then Eq.~\ref{eq:Maxwell} becomes
\begin{equation}
\frac{\partial^{2} \psi}{\partial x^{2}}+(1+\frac{\omega_{M}\omega_{T}}{\omega_{T}\omega_{x}-\omega^{2}})\frac{\partial^{2} \psi}{\partial y^{2}}+(1+\frac{\omega_{M}\omega_{x}}{\omega_{T}\omega_{x}-\omega^{2}})\frac{\partial^{2} \psi}{\partial z^{2}}=0.
\end{equation}
Recalling that
\begin{equation}
\psi = a e^{i(\bm{k}\cdot\bm{r}-\omega t)},
\end{equation}
we have
\begin{equation}
\label{eq:Laplacian}
k_{x}^{2}+(1+\frac{\omega_{M}\omega_{T}}{\omega_{T}\omega_{x}-\omega^{2}})k_{y}^{2}+(1+\frac{\omega_{M}\omega_{x}}{\omega_{T}\omega_{x}-\omega^{2}})k_{z}^{2}=0.
\end{equation}
By solving Eq.~\ref{eq:Laplacian}, we get the dispersion relation of the dipolar magnon mode,
\begin{equation}
\omega = \frac{\sqrt{\omega_{T}\omega_{x}k_{x}^{2}+\omega_{T}(\omega_{M}+\omega_{x})k_{y}^{2}+\omega_{x}(\omega_{M}+\omega_{T})k_{z}^{2}}}{\sqrt{k_{x}^{2}+k_{y}^{2}+k_{z}^{2}}}
\end{equation}

\begin{figure} [ht!]
\includegraphics[width=5.5in]{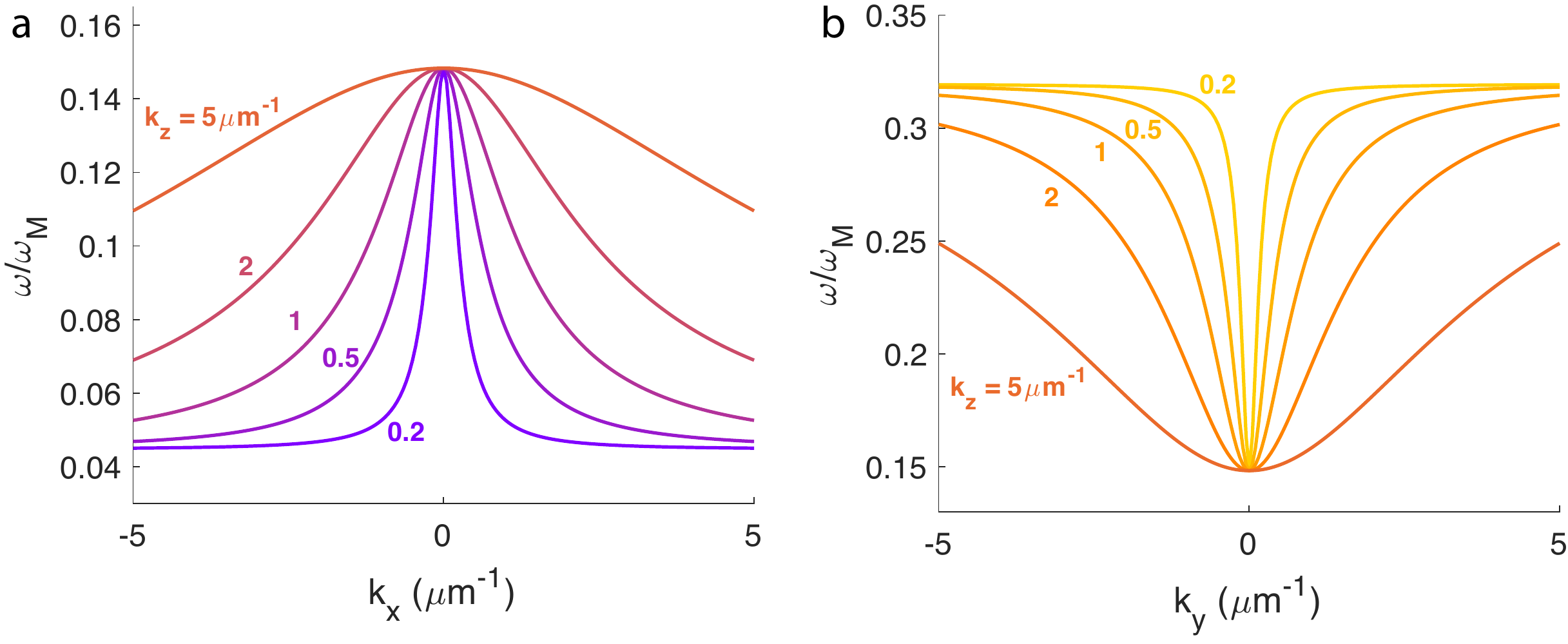}
\caption{\label{figS2} The $k_z$ dependence of the dipolar magnon dispersion. Line cuts taken through the plane defined by (a) $k_y = 0$ and (b) $k_x = 0$ are shown for selected values of $k_z$.}
\end{figure}

The dipolar magnon dispersion relations at selected values of $k_z$ are plotted in Fig. \ref{figS2} using the following parameters: $\omega_x = 4.24 $ rad/s, $\omega_T = 21.18 $ rad/s, and $\omega_M = 211.81$ rad/s. The magnon bands become more dispersive for smaller values of $k_z$.

\section{The dipolar magnon dispersion in an external field}
We now include an external field applied along the direction of the hard axis ($z$-axis). The free energy becomes,
\begin{equation}
F=-K_{x}\frac{M_{x}^{2}}{M_{s}^{2}}+K_{z}\frac{M_{z}^{2}}{M_{s}^{2}}-H_{z}M_{z}.
\end{equation}
By minimizing the free energy, we show that the equilibrium magnetization in an applied field $H_{z}$ is $\bm{M}=(M_{s}\cos\theta,0,M_{s}\sin\theta)$, where
\begin{equation}
\theta=
    \begin{dcases}
        \arcsin\left(\frac{H_{z}}{H_{s}}\right), & 0\leqslant H_{z} \leqslant H_{s}\\
        \frac{\pi}{2}, & H_{z}>H_{s}
    \end{dcases}
\end{equation}
$H_{s}$ is the saturation field, and $H_{s} = 2(K_{x}+K_{z})/M_{s}$. The normal mode of magnons is a precession in the plane orthogonal to the equilibrium magnetization, so it has the form $\bm{m}=(-m_{a}\sin\theta,m_{y},m_{a}\cos\theta)$. Following the similar approach in Sec. \ref{zero_field}, we have the magnon dispersion below.

For $0\leqslant H_{z} \leqslant H_{s}$,
\begin{equation}
\begin{aligned}
&\omega = \frac{\gamma}{K_{x}+K_{z}}\frac{\sqrt{c_{x}k_{x}^{2}+c_{y}k_{y}^{2}+c_{z}k_{z}^{2}-c_{xz}k_{x}k_{z}}}{\sqrt{k_{x}^{2}+k_{y}^{2}+k_{z}^{2}}},
\\& c_{x} = K_{x}[(H_{s}^{2}-H_{z}^{2})(K_{x}+K_{z})+2\pi H_{z}^{2}M_{s}^{2}],
\\& c_{y} = (H_{s}^{2}-H_{z}^{2})(K_{x}+2\pi M_{s}^{2})(K_{x}+K_{z}),
\\& c_{z} = (H_{s}^{2}-H_{z}^{2})K_{x}(K_{x}+K_{z}+2\pi M_{s}^{2}),
\\& c_{xz} = H_{z}\sqrt{H_{s}^{2}-H_{z}^{2}}K_{x}M_{s}^{2}.
\end{aligned}
\end{equation}

For $H_{z}>H_{s}$,
\begin{equation}
\begin{aligned}
&\omega = \gamma\frac{\sqrt{d_{x}k_{x}^{2}+d_{y}k_{y}^{2}+d_{z}k_{z}^{2}}}{\sqrt{k_{x}^{2}+k_{y}^{2}+k_{z}^{2}}},
\\& d_{x} = \left(H_{z}-\frac{2K_{z}}{M_{s}}\right)(H_{z}-H_{s}+4\pi M_{s}),
\\& d_{y} = (H_{z}-H_{s})\left(H_{z}-\frac{2K_{z}}{M_{s}}+4\pi M_{s}\right),
\\& d_{z} = \left(H_{z}-\frac{2K_{z}}{M_{s}}\right)(H_{z}-H_{s}).
\end{aligned}
\end{equation}

When $\bm{k}\to 0$, we get the $k=0$ frequency as a function of the applied field,
\begin{equation}
\omega(k=0) =
    \begin{dcases*}
        \frac{\gamma\sqrt{(H_{s}^{2}-H_{z}^{2})K_{x}(K_{x}+K_{z}+2\pi M_{s}^{2})}}{K_{x}+K_{z}}, & $0\leqslant H_{z} \leqslant H_{s}$ \\
        \gamma\sqrt{\left(H_{z}-\frac{2K_{z}}{M_{s}}\right)(H_{z}-H_{s})},, & $H_{z}>H_{s}$
    \end{dcases*}
\end{equation}

\section{Micromagnetic Simulations}
The micromagnetic simulations are performed using the program MuMax3. The purpose of the micromagnetic simulation is to numerically calculate the magnon dispersion of Fe$_{3}$Sn$_{2}$ based on a time evolution simulation.

In the simulation, the sample size is set to be 50 $\mu$m (length) $\times$ 50 $\mu$m (width) $\times$ 3.125 $\mu$m (thickness) and the micromagnetic cell size is 0.098 $\mu$m $\times$ 0.098 $\mu$m $\times$ 0.098 $\mu$m. The periodic boundary condition is applied along the x direction. The simulation is performed by setting the saturation magnetization $M_{s}=6.495\times10^{5}$ A/m, the exchange constant $A_{ex}=1.4\times 10^{-11}$ J/m, the hard-axis anisotropy along z-axis $K_{z}=2.317\times 10^{5}$ J/m$^{3}$ and the easy-axis anisotropy along x-axis $K_{x}=1.186\times 10^{4}$ J/m$^{3}$.

An oscillating external magnetic field along z direction is applied to excite the magnons. To obtain a clear picture of the magnon dispersion, the excitation of the time evolution needs to cover as many wavevectors as possible. Therefore, the spatial profile of the excitation is localized at the very center micromagnetic cell in each xy layer, so that the broadest distribution in k-space can be achieved. The temporal profile of the excitation is a narrow sinc pulse
\begin{equation}
    B_{z} = 0.01\sinc[2\pi f(t-t_{0})],
\end{equation}
where $f=50$ GHz and the total simulation time $t_{0}=10$ ns. This temporal profile yields a uniform window in the frequency domain to observe the magnon dispersion.

\begin{figure} [ht!]
\includegraphics[width=5.5in]{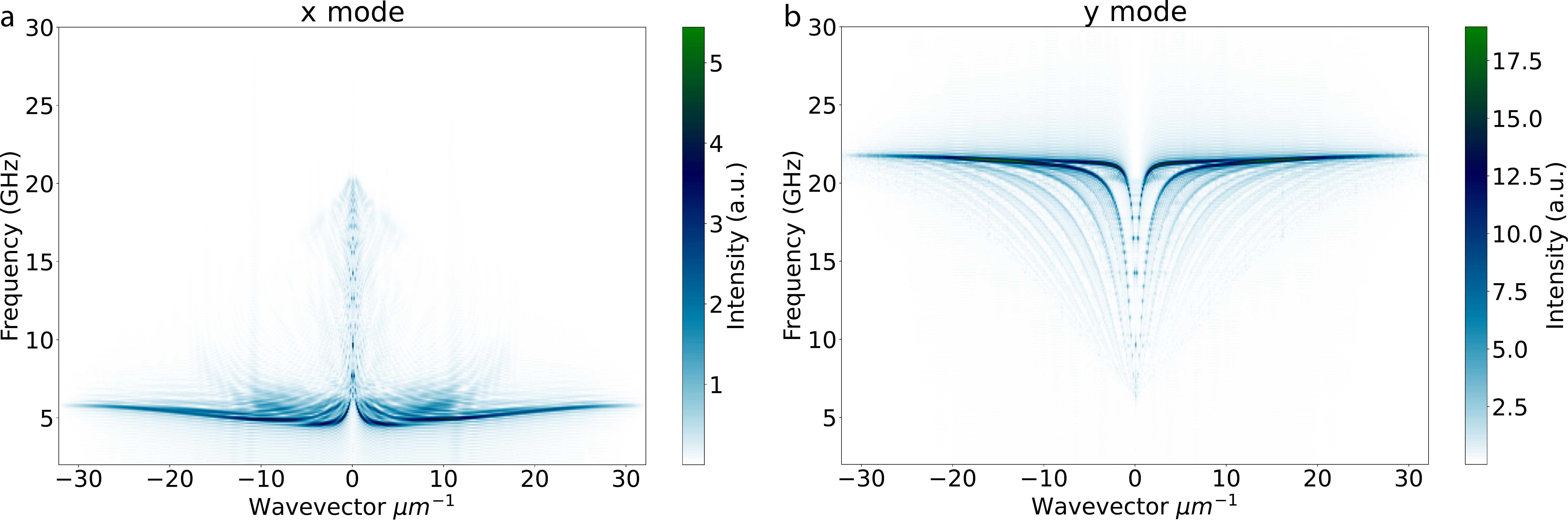}
\caption{Micromagnetic simulations. The magnon dispersion relations calculated for a biaxial ferromagnet show (a) backward propagating and (b) forward propagating bulk modes, consistent with the calculations described in Section \ref{zero_field}.}\label{figS3} 
\end{figure}

After obtaining the time evolution of each micromagnetic cell, we extract the middle $xy$ layer of the sample and perform a 3D Fourier transform with respect to $x$, $y$ and $t$. The Fourier amplitude gives the frequency $\omega$ as a function of $k_{x}$ and $k_{y}$, or equivalently the magnon dispersion relation. The simulation results are shown in Fig. \ref{figS3}.

\section{Numerical simulations}
\begin{figure} [ht!]
\includegraphics[width=4.5in]{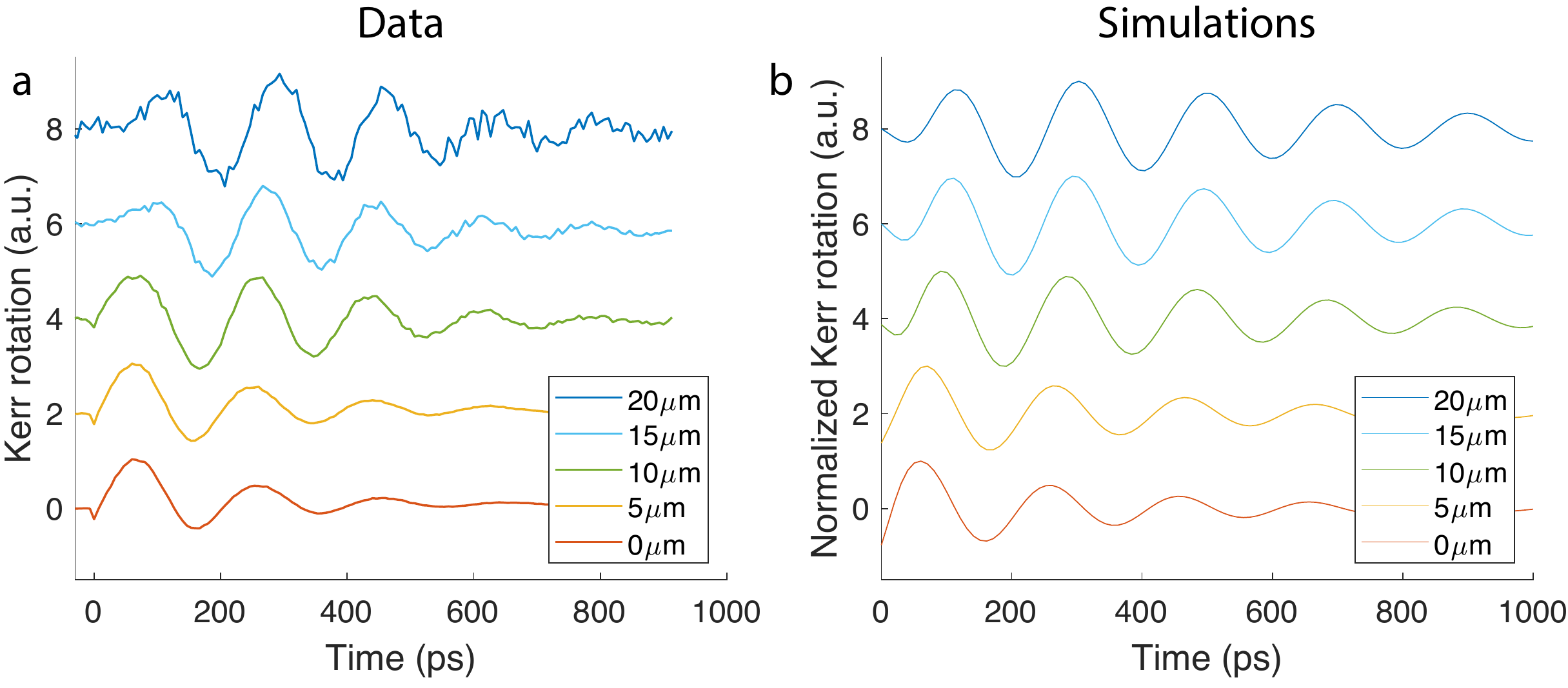}
\caption{\label{figS4} Numerical simulations. (a) Experimental TR-MOKE data traces at selected values of spatial separation $\Delta x$ between the pump and probe pulses. (b) Numerically simulated TR-MOKE traces chosen at the same values $\Delta x$ show good agreement with the data}
\end{figure}

Numerical simulations are carried out using the following expressions (Eq. (3) of the main text):
\begin{align}
    \delta {M_z}(\textbf{r},t) \propto\textrm{Re}\iiint \textbf{z}\cdot\textbf{m}(\textbf{k})g(\textbf{k})e^{i\textbf{k} \cdot \textbf{r}} e^{-i[\omega(\textbf{k})-i\alpha ]t}dk_x dk_y dk_z,
\end{align}
with
\begin{align}
    g(\textbf{k})\propto\frac{e^{-\sigma^2(k_x^2+k_y^2)/2} }{ik_z - 1/\delta_p},
\end{align}
which describes time and spatial evolution of the magnetization vector after its transient misalignment from the effective anisotropy field direction. The integration bounds were set to $-10/d < k_{x,y} < 10/d$ and $0.01/\delta_p < |k_z| < 3/\delta_p$, where $d$ is the full width at half maximum spot size of the pump beam and $\delta_p$ is the effective penetration depth. A Gaussian smoothing was applied to the calculated values of $M_z(r,t)$ to take into account the finite spot size of the probe focus (FWHM $=5 \mu$m). The best fit to the data was found by setting $\delta_p = 230$ nm and  $\alpha = 370 $ ps, which are the only fitting parameters of the simulations. The time-resolved traces acquired from numerical simulations are in good agreement with the experimental data, as shown in Fig. \ref{figS4}.

\bibstyle{apsrev4-2}
\bibliography{bibliography_SI}